\documentclass{IEEEtran}
\usepackage{cite}
\usepackage{amsmath,amssymb,amsfonts}
\usepackage{algorithmic}
\usepackage{graphicx}
\usepackage{textcomp}
\usepackage{array}
\usepackage[utf8]{inputenc}
\usepackage{setspace}
\def\BibTeX{{\rm B\kern-.05em{\sc i\kern-.025em b}\kern-.08em
    T\kern-.1667em\lower.7ex\hbox{E}\kern-.125emX}}
\begin{document}
\title{Ultra-Wideband Antenna with MIMO Diversity for 5G Wireless Communication}

\author{M. Saeed Khan, Adnan Iftikhar, Raed M. Shubair, Antonio-D. Capobianco, Benjamin D. Braaten,  and Dimitris E. Anagnostou

\thanks{A. Iftikhar is with the Electrical Engineering Department, COMSATS University, Islamabad,
Islamabad, 45550, Pakistan e-mail: (adnaniftikhar@comsats.edu.pk.}
\thanks{R. M. Shubair is with Massachusetts Institute of Technology (MIT), Cambridge, MA, USA and New York University, Abu Dhabi.}
\thanks{A. -D. Capobianco is with the Department of Information Engineering, University of Padova, Via Gradenigo 6/b, 35131,
Padova, Italy.}
\thanks{B. D. Braaten is with the Department of Electrical and Computer Engineering, North Dakota State University, Fargo, ND 58102, USA.}
\thanks{D. E. Anagnostou is with the Department of Electrical Engineering, Heriot-Watt University,
Edinburgh, U.K.}}

\maketitle

\begin{abstract}
An eight element, compact Ultra Wideband$-$ Multiple Input Multiple Output (UWB-MIMO) antenna capable of providing high data rates for future Fifth Generation (5G) terminal equipments along with the provision of necessary bandwidth for Third Generation (3G) and Fourth Generation (4G) communications that accomplishes band rejection from 4.85 to 6.35 GHz by deploying a Inductor Capacitor (LC) stub on the ground plane is presented. The incorporated stub also provides flexibility to reject any selected band as well as bandwidth control. The orthogonal placement of the printed monopoles permits polarization diversity and provides high isolation. In the proposed eight element UWB-MIMO/diversity antenna, monopole pair 3$-$4 are 180$^{\circ}$ mirrored transform of monopole pair 1$-$2 which lie on the opposite corners of a planar 50~\texttimes~50 mm$^{2}$ substrate. Four additional monopoles are then placed perpendicularly to the same board leading to a total size of 50~\texttimes~50~\texttimes~25 mm$^{3}$ only. The simulated results are validated by comparing the measurements of a fabricated prototype. It was concluded that the design meets the target specifications over the entire bandwidth of 2 to 12 GHz with a reflection coefficient better than $-$10 dB (except the rejected band), isolation more than 17 dB, low envelope correlation, low gain variation, stable radiation pattern, and strong rejection of the signals in the Wireless Local Area Network (WLAN) band. Overall, compact and reduced complexity of the proposed eight element architecture, strengthens its practical viability for the diversity applications in future 5G terminal equipments amongst other MIMO antennas designs present in the literature.
\end{abstract}

\begin{IEEEkeywords}
Band rejected, compact, diversity, envelope correlation co-efficient, multiple input multiple output, ultrawide band, 5G communication, 5G terminal devices.
\end{IEEEkeywords}

\clearpage

\section{Introduction}
\label{sec:introduction}
\IEEEPARstart{W}{ireless} broadband communication system such as Worldwide Interoperability for Microwave Access (WiMAX (3.4 to 3.6 GHz)), large capacity Microwave Relay Trunk Network (4.4 to 4.99 GHz), and Wireless Local Area Network (WLAN signals in 5.15 to 5.35 and 5.75 to 5.8225 GHz bands), impose a limited power spectral density of the low power and a high data rate UWB signal, even though, the bandwidth of UWB is wide (3.1 GHz – 10.6 GHz) \cite{kaiser2009overview}. The amalgamation of UWB technology with MIMO permits such wireless systems to achieve high data rates by transmitting wireless signals over multiple channels without increasing the input power. For instance, high speed and data rates in Wireless Personal Area Networks (WPAN) is only possible with UWB-MIMO technology \cite{wallace2003experimental}. One key specification of such MIMO antennas is that isolation between its elements should be more than 15 dB to ensure mitigation of inter-element Electromagnetic Interference (EMI) \cite{satam2018spanner,liu2013compact,khan2015planar,anitha2016compact,lin2009ultra}. The inter-element EMI can be mitigated by ensuring $\lambda$/2 spacing between the MIMO antenna elements. However, the spacing significantly effects the compactness of the MIMO antennas which ultimately employ practical restrictions on the MIMO antenna to be integrated in indoor and outdoor portable devices. Besides, researchers have proposed various techniques to mitigate inter-element EMI without effecting the compactness of the MIMO antenna \cite{satam2018spanner,liu2013compact,khan2015planar,anitha2016compact,lin2009ultra,khan20154,khan2016compact,kiem2014design,wu2018quad,knudsen2002spherical,sharawi20135,al2014eight,saleem2015eight,li2017eight,sipal2017easily}. In addition, other wireless communication standards such as WiMAX, WLAN, and X-band downlink frequencies may electromagnetically interfere with the UWB spectrum and may affect systems' performance. Therefore, several techniques have been reported in literature to mitigate the EMI from these wireless communication standards that are allocated in the UWB spectrum.

A high gain two-element spanner shaped UWB-MIMO with edge truncation (for bandwidth enhancement) is proposed for UWB applications \cite{satam2018spanner}. The two element UWB-MIMO antenna provides an isolation higher than 15 dB by placing elements 0.25 $\lambda$ apart, but, expansion of the proposed antenna to eight element MIMO antenna significantly enlarges overall size. Stubs have been employed on the ground plane between two orthogonally placed elements to improve the isolation \cite{liu2013compact}. Another design in \cite{khan2015planar}, also exploits the polarization diversity for UWB-MIMO applications. A four element design with modified slotted ground plane is presented in \cite{anitha2016compact}. The proposed solution reaches an isolation of 14 dB with small (45~\texttimes~45 mm$^{2}$) size over the bandwidth of 2 to 6 GHz. Another design in \cite{lin2009ultra}, exhibits an isolation of more than 20 dB with discontinuities between the antenna elements and ground plane, however the size of the antenna becomes too large (110~\texttimes~114 mm$^{2}$) for 2 to 6 GHz band. Compact designs (with size 39.8~\texttimes~50 mm$^{2}$) have been proposed for four element UWB-MIMO application in \cite{khan20154,khan2016compact}. The elements exploit the polarization diversity to obtain isolation of more than 17 dB with an addition of complex band stop design on the back side of the radiators to reject the WLAN band in \cite{khan2016compact}. With a total size of 60~\texttimes~60 mm$^{2}$, an electromagnetic Bandgap (EBG) structure is employed in \cite{kiem2014design,wu2018quad} to reject the WLAN for four element UWB-MIMO antenna. Mushroom like stub structure is used in \cite{kiem2014design} for obtaining an isolation higher than 15 dB, while polarization diversity is exploited in \cite{wu2018quad} for an isolation of more than 17.5 dB, also large number of vias are used for band rejection. In \cite{sharawi20135}, an eight element array with complementary split$-$ring resonators (CSRR) is detailed to achieve 20 dB isolation in MIMO system. In another design, array elements are placed at large distance in eight element MIMO array to obtain an isolation of 10 dB \cite{al2014eight}. In \cite{saleem2015eight}, an eight element planar antenna is proposed for UWB-MIMO applications with a complex structure on the bottom side of the elements and large dimensions of the board. Narrow band polarization diversity antenna with eight elements is proposed for fifth$-$generation (5G) application in \cite{li2017eight}. Identical elements over isolated ground plane are used to obtain high isolation in an eight port UWB-MIMO design \cite{sipal2017easily}, while polarization and pattern diversity is investigated in \cite{mathur20198} by deploying two different slots. Open slot metal frame is used to design eight port UWB-MIMO antenna for 5G communications \cite{zhang2019ultra}. Dual notch band, sharp rejection of narrow band, and reconfigurability of the notch band using RF$-$MEMS has been explored on a single UWB radiator and it has been proven that rejection quality is inversely proportional to the rejection band \cite{anagnostou2007dual,gheethan2012dual,anagnostou2013reconfigurable}.

All the aforementioned UWB-MIMO designs have tradeoffs between design complexity, size, number of ports and bandwidth. The design reported in \cite{satam2018spanner,liu2013compact,khan2015planar,anitha2016compact,lin2009ultra,khan20154,khan2016compact,kiem2014design} are suitable for two ports or four ports in planar configuration and no guidelines are presented to extend the design. Also, if the design is extended for a large number of ports, it becomes too large to be deployed in practical scenarios. In most of the eight elements cases, the design compactness exceeds 0.25 $\lambda$ \cite{saleem2015eight,sipal2017easily,mathur20198}, and has a narrow bandwidth \cite{sharawi20135,al2014eight,li2017eight,zhang2019ultra}. Last but not the least, all the designs proposed with the eight elements have not band rejection capabilities. Whereas, the proposed design, when compared to current designs, have a clear advantage of compactness and tunable capability to reject various bands with in the UWB spectrum.

Moreover, conventional 2~\texttimes~2 or 4~\texttimes~4 UWB-MIMO antenna for future 5G Customer Premises Equipment (CPE) may not meet high data rates as required by the 5G technology. The current deployment of 5G technology in developed countries (USA, Japan, and China) utilizes sub$-$6 GHz band i.e. LTE band 42 (3.4 GHz$-$3.6 GHz) and LTE band 43 (3.6 GHz$-$3.8 GHz). Besides, non-standalone solutions with 4G as an anchor, deployed in Gulf Corporation Council (GCC) countries for 5G communication operates on sub$-$6 GHz (3.5 GHz) band having operational bandwidth of 50 MHz$-$100 MHz. To achieve higher data rates in the 5G technology, some 8 $\times$ 8 MIMO \cite{ban20164g,li2016eight,guo2018side,li201712}, 8~\texttimes~8 UWB-MIMO \cite{zhang2019ultra}, and 10~\texttimes~10 MIMO \cite{li2018multiband} antennas have been proposed for the future 5G devices and those can be utilized for the 5G CPE such as routers etc. However, the planar configuration of the reported design along with the incapability to provide necessary bandwidth for 4G and 3G communications having band rejection feature may not be an appropriate solution for the forthcoming 5G technology because of low footprint requirements. Likewise, four element reconfigurable band reject UWB-MIMO antenna, having two elements in planar configuration and two elements fixed at $\pm$ 45$^\circ$ is proposed in \cite{khan2020ultra}. The proposed configuration offers excellent band rejection, however, hardware complexity and fixation of angularly placed elements are major hindrances of this design for modern 4G/5G communication devices. In addition, a 3-D eight element UWB-MIMO array is presented in \cite{khan2018compact}, but without any band rejection capabilities. The addition of band rejection structures in the existing 3D UWB-MIMO may affect impedance, isolation, and radiation characteristics and result in complex geometrical configurations \cite{khan2018compact}.

Therefore, the objective and novelty of this work is; (a) utilize the space provided for the antenna more efficiently by incorporating as many elements as possible for future 5G technology, without increasing the size of the board, by placing additional elements perpendicular to the board, (b) introduce the WLAN band rejection capability in all elements without affecting the performance of the other elements. Keeping in mind all of the above aspects, an eight element UWB-MIMO/diversity antenna with WLAN band rejection capabilities is proposed here. The band rejection capabilities obtained by quarter wavelength stub on the ground plane can also be used to reject other bands by modifying the length of the stub. Four monopoles exploiting the polarization diversity amongst them are placed in a planar configuration on a 50~\texttimes~50 mm$^{2}$ board and four additional monopoles are then adjusted perpendicularly to the same board in order to obtain a compact size. The orthogonal polarization from the closely spaced monopoles guarantees high inter-element isolation.

The work of this project adds to the previous contributions \cite{7696590,7928624,8231103,7305541,6512176,7696590,8888295,8292726,8530987,khan2016compact,khan2016compact1,khan2017ultra,omar2016uwb,shubair2015novel,shubair2015novel1,al2006direction,al2005direction,nwalozie2013simple,shubair2005robust,belhoul2003modelling,ibrahim2017compact,shubair2004robust,che2008propagation,el2016design,shubair1993closed,al2005computationally,al2003investigation,shubair2005performance,al2016millimeter,shubair2005convergence,khan2018triband,alharbi2018flexible,ibrahim2016reconfigurable,hakam2016novel,shubair1992simple,ghosal2018characteristic,khan2018novel,elsalamouny2015novel,khan2016pattern,shubair2005improved,8458184,7156466,7696430,7347957,8646600,7305294,8706527,7803806,7803808,9082193,alhajri2018accurate} reported in the literature in new applications and evolving technologies.

\clearpage

\section{Design Procedure}
 Initially, a wideband rectangular monopole with arc$-$shaped feeding section and a partial ground plane was designed in 3D EM software, ANSYS HFSS. The ground width ‘w’ is so chosen to obtain good impedance matching throughout the entire frequency range. For further optimization, corners were truncated for impedance enhancement at higher frequencies and impedance transformer was attached to the tapered arc section for impedance enhancement at lower frequencies. After achieving the impedance matching over the wide bandwidth (2 to 12 GHz), a stub was connected to the ground on the backside of the monopole to reject the WLAN band. The length of the stub was calculated using $L_{f}=\dfrac{\lambda_{g}}{4}=\dfrac{c}{4f_{\circ}{\sqrt{\epsilon_{r}}}}$ where $f_{\circ}~\approx$ 5 GHz, and $\epsilon_{r}$ is the relative permittivity of the substrate. The length can also be varied to reject other bands by changing $f_{\circ}$. The position of the stub to connect the ground plane was chosen after carefully observations of surface currents on the ground plane at 5.8 GHz and it was found that centre of the ground plane beneath the transmission line is the most effective place for the stub connection to draw the current on the stub.
 \begin{figure*}[!t]
\centerline{\includegraphics[width=7 in]{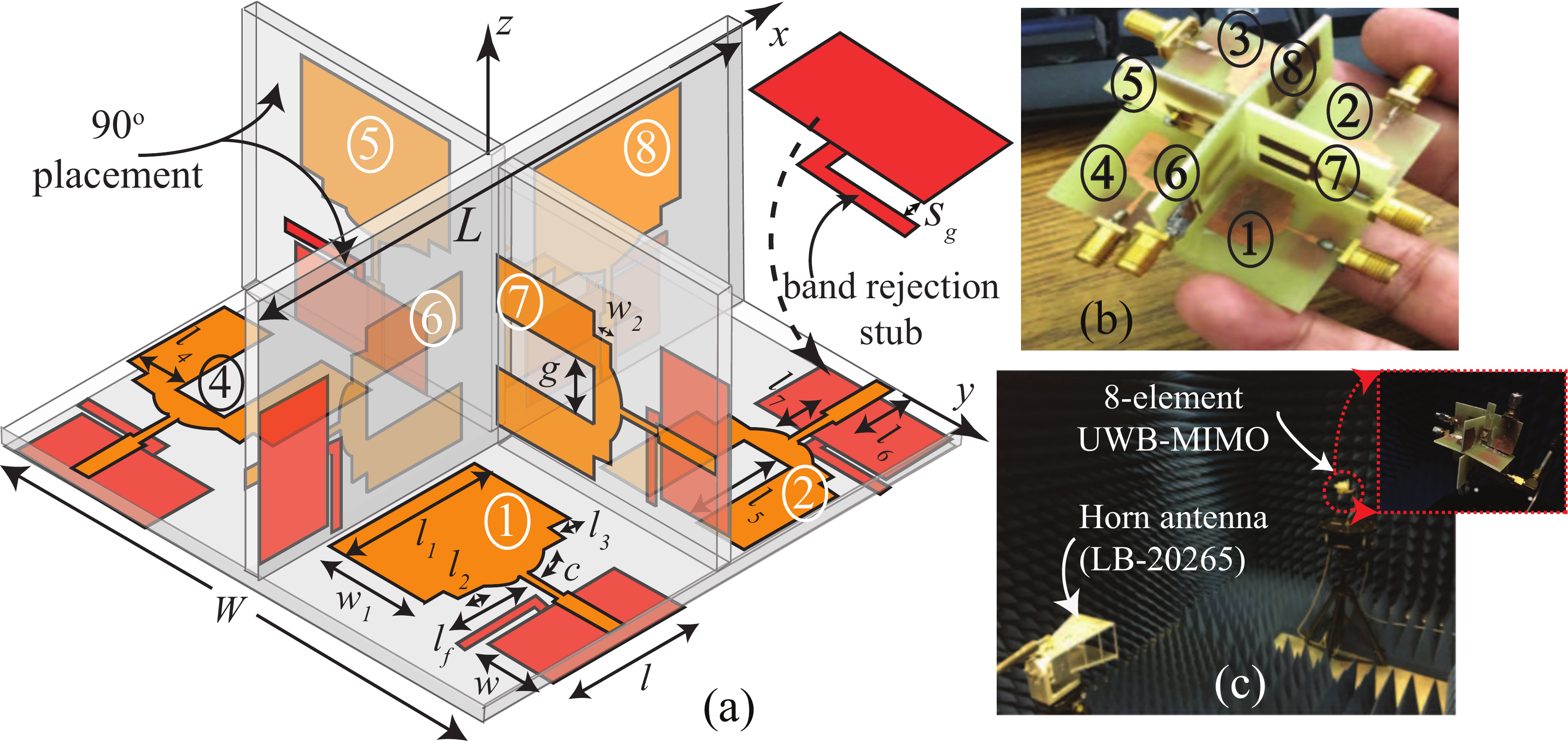}}
\caption{(a) Perspective view of the proposed UWB-MIMO antenna with dimension in mm. $L$ = 50, $W$ = 50, $w_{1}$ = 10, $l_{1}$ = 15, $l_{2}$ = 2.25, $l_{3}$ = 2, $l_{4}$ = 5, $l_{5}$ = 10, $g$ = 5, $w_{2}$ = 1.5, $l_{6}$ = 6, $l_{7}$ = 3.82, $c$ = 3.1, $l$ = 13.5, $w$ = 7, $L_{f}$ = 7.25, and $S_{g}$ = 0.5, (b) perspective view of fabricated prototype with dimensions shown in Fig. 1, and (c) photo during pattern measurement in fully calibrated anechoic chamber.}
\label{fig1a}
\end{figure*}

\begin{figure*}[!h]
\centerline{\includegraphics[width= 7 in]{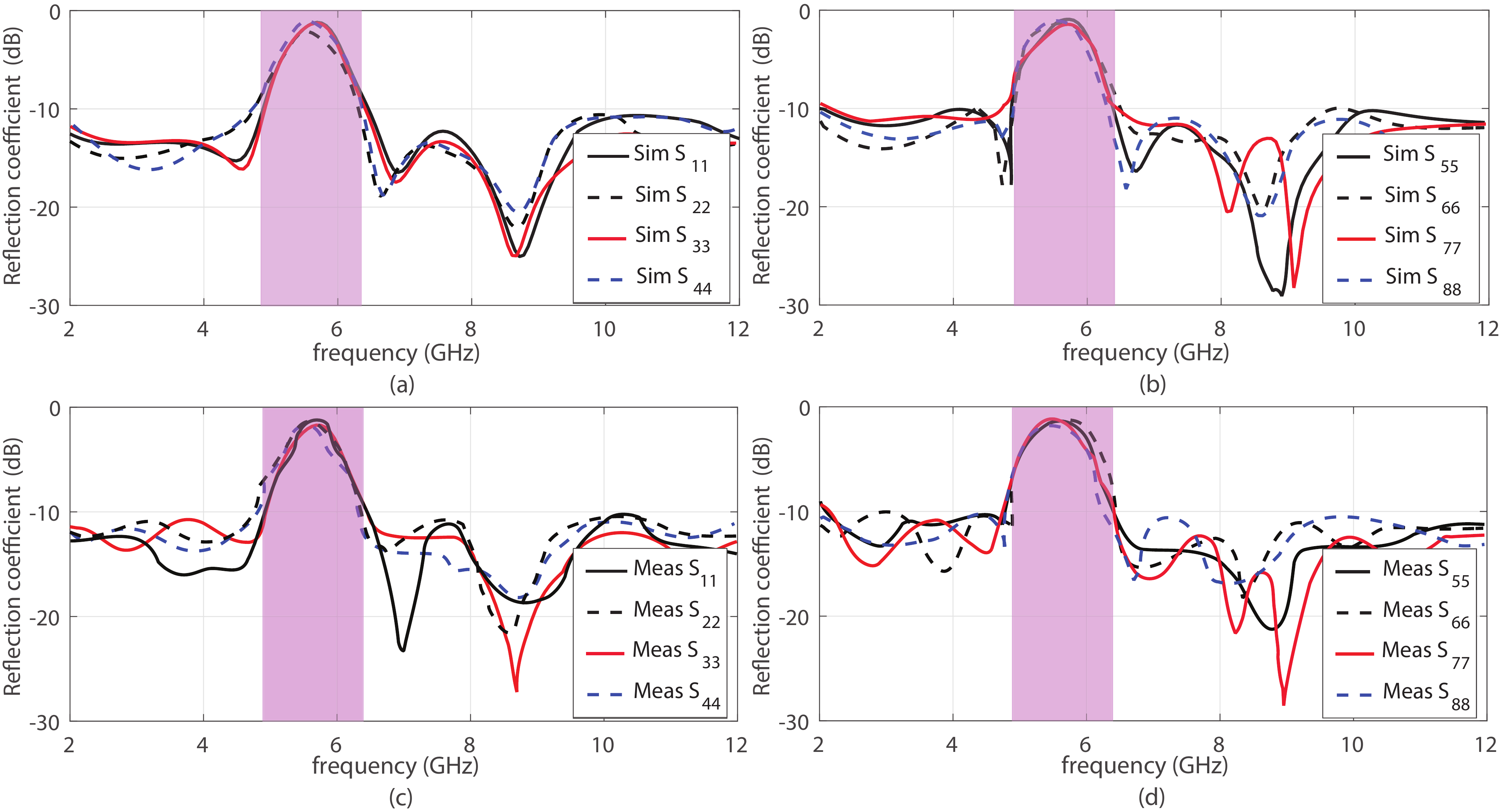}}
\caption{Simulated and measured reflection coefficient below $-$10 dB except the rejected band, where values reach $-$1 dB almost at all input ports, (a) simulated port 1 to 4, (b) simulated port 5 to 8, (c) measured port 1 to 4, and (d) measured port 5 to 8.}
\label{fig1b}
\end{figure*}

 After obtaining the band rejection capabilities for a single monopole, a second one was placed orthogonally at 6.15 mm edge distance to exploit the polarization diversity. The distance was chosen after parametric study and to have space for the inclusion of more elements later on. The placement of second monopole induced surface currents and affected the impedance match of the second monopole at lower frequencies, which was improved by inserting a U$-$shaped slot \cite{satam2018spanner}. The polarization purity of element 1 is high while the polarization of element 2 is low. This is  because  of  the  U-shaped  slot  which  decreased the power level of antenna element 2. The parameters of the slots were optimized using $L_{5}=\dfrac{\lambda_{g}}{4}=\dfrac{c}{4f_{\circ}{\sqrt{\epsilon_{r}}}}$ and $g=\dfrac{\lambda_{g}}{8}=\dfrac{c}{8f_{\circ}{\sqrt{\epsilon_{r}}}}$ where $f_{\circ}~\approx$ 3.60 GHz, and $\epsilon_{r}$ is the relative permittivity of the substrate. Two more elements were then placed in planar configuration on a 50 $\times$ 50 mm$^{2}$ area. The placement of the monopoles was numerically adjusted to obtain low mutual coupling and impedance matching over the entire bandwidth. The edge distance between element 1 and 4 was kept 12 mm so that the design can be placed at the edge of the PCB board and deliver high isolation. After obtaining the desired performance of four monopoles in the planar configuration, four more monopoles (labelled as 5, 6, 7 and 8 in Fig. 1 (a)) were employed perpendicularly, preserving the compactness of the design. The polarization diversity was also achieved in these monopoles along with the wideband impedance matching, low mutual coupling, and band rejection capabilities.
 The dimensions of the proposed 3$-$D eight elements UWB-MIMO array (total volume 50~\texttimes~50~\texttimes~25 mm$^{3}$) are shown in Fig. 1 (a). In the proposed design configuration, the separated ground plane is used to suppress the surface current and mitigate the near-field coupling. Since antenna elements are placed too closely it is necessary to disconnect the ground plane or use any complex decoupling structure. In the proposed configuration, disconnected ground plane was preferred, because four elements (elements 5 to 8) were placed in vertical configuration.

\begin{figure*}[!t]
\centerline{\includegraphics[width= 6 in]{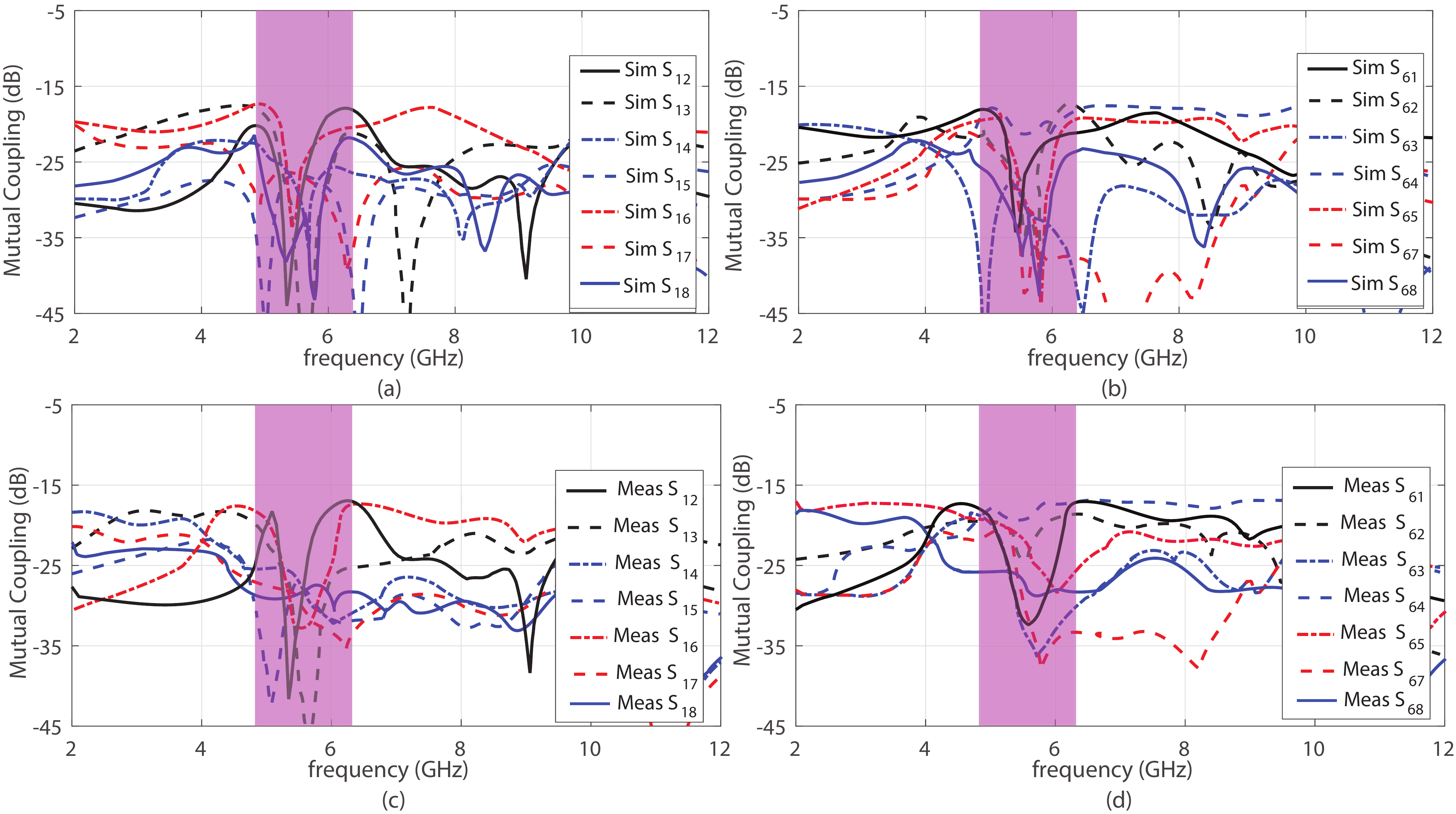}}
\caption{Simulated and measured mutual couplings less than $-$17 dB amongst all ports, (a) simulated port 1, (b) measured port 6, (c) measured port 1, and (d) measured port 6.}
\label{fig1c}
\end{figure*}

\begin{figure*}[!h]
\centerline{\includegraphics[width= 6 in]{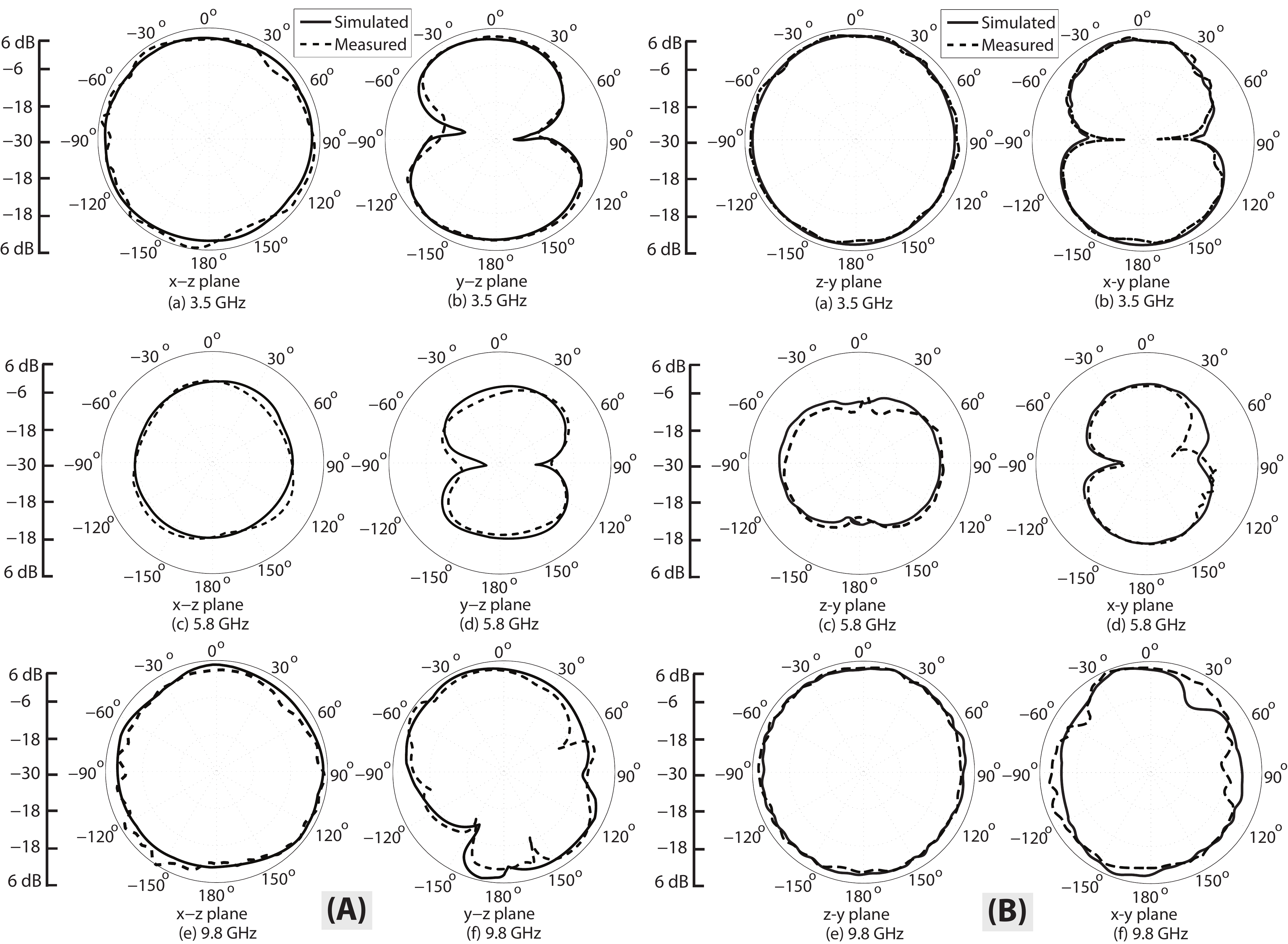}}
\caption{(A) Simulated and measured radiation patterns for port 1 in the principle planes, (a) x$-$z plane at 3.5 GHz, (b) y$-$z plane at 3.5 GHz, (c) x$-$z plane at 5.8 GHz, (d) y$-$z plane at 5.8 GHz, (e) x$-$z plane at 9.8 GHz, and (f) y$-$z plane at 9.8 GHz. The patterns are nearly omni$-$directional in the x$-$z plane, suitable for UWB-MIMO systems and (B) simulated and measured radiation patterns for port 6 in the principle planes, (a) z$-$y plane at 3.5 GHz, (b) x$-$y plane at 3.5 GHz, (c) z$-$y plane at 5.8 GHz, (d) x$-$y plane at 5.8 GHz, (e) z$-$y plane at 9.8 GHz, and (f) x$-$y plane at 9.8 GHz. The patterns are nearly omni$-$directional in the z$-$y plane, suitable for UWB-MIMO systems.}
\label{fig1d}
\end{figure*}

\clearpage

\section{Results and Discussion}
\subsection{S$-$parameters}
The layout shown in Fig. 1 (a) was printed on a FR4 board (thickness = 1.6 mm, $\epsilon_{r}$ = 4.5 and tan$\delta$ = 0.02), as shown in Fig. 1 (b). The prototype was then characterized using an Agilent N5242A PNA$-$X network analyser. Fig. 2 (a) to (d) shows the simulated and measured reflection coefficients at all input ports. The measured reflection coefficients were lower than $-$10 dB at all ports except in the rejected band (4.85 to 6.35 GHz). Good agreement between simulated and measured results was found, with a slight variation ($\pm$ 1.1 dB in most of the band) due to manufacturing imperfections. The simulated and measured mutual couplings of the ports 1 and 6 are also plotted in Fig. 3. For measurement purposes, the ports 1 and 6 are selected to realize the effect on the monopole in planar configuration and perpendicular placement of the monopoles. It can be seen in Fig. 3 (a) to (d) that both simulated and measured mutual couplings are not exceeding $-$17 dB level, which is a significant achievement for having such a large number of antennas in such a compact volume.

\begin{figure}[!t]
\centerline{\includegraphics[width= 2.5 in]{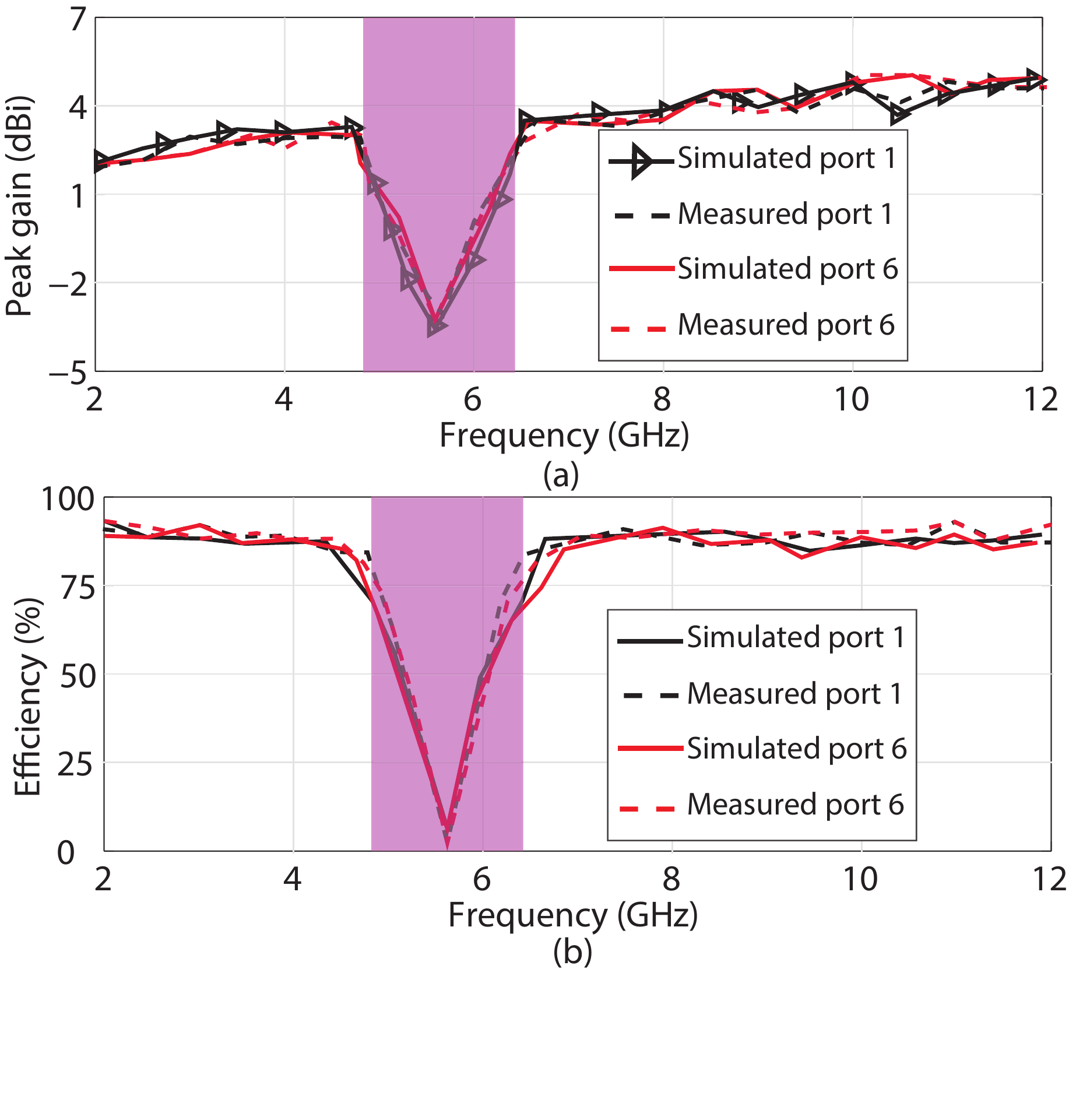}}
\caption{(a) Simulated and measured peak gain over the entire band, the gain varies from 2.65 dBi to 5.8 dBi except the rejected band, where gain drops to -3.6 dBi. (b) Simulated and measured efficiency over 86 $\%$ in the entire band except the rejected band.}
\label{figgg}
\end{figure}

\begin{figure}[!h]
\centerline{\includegraphics[width= 2.5 in]{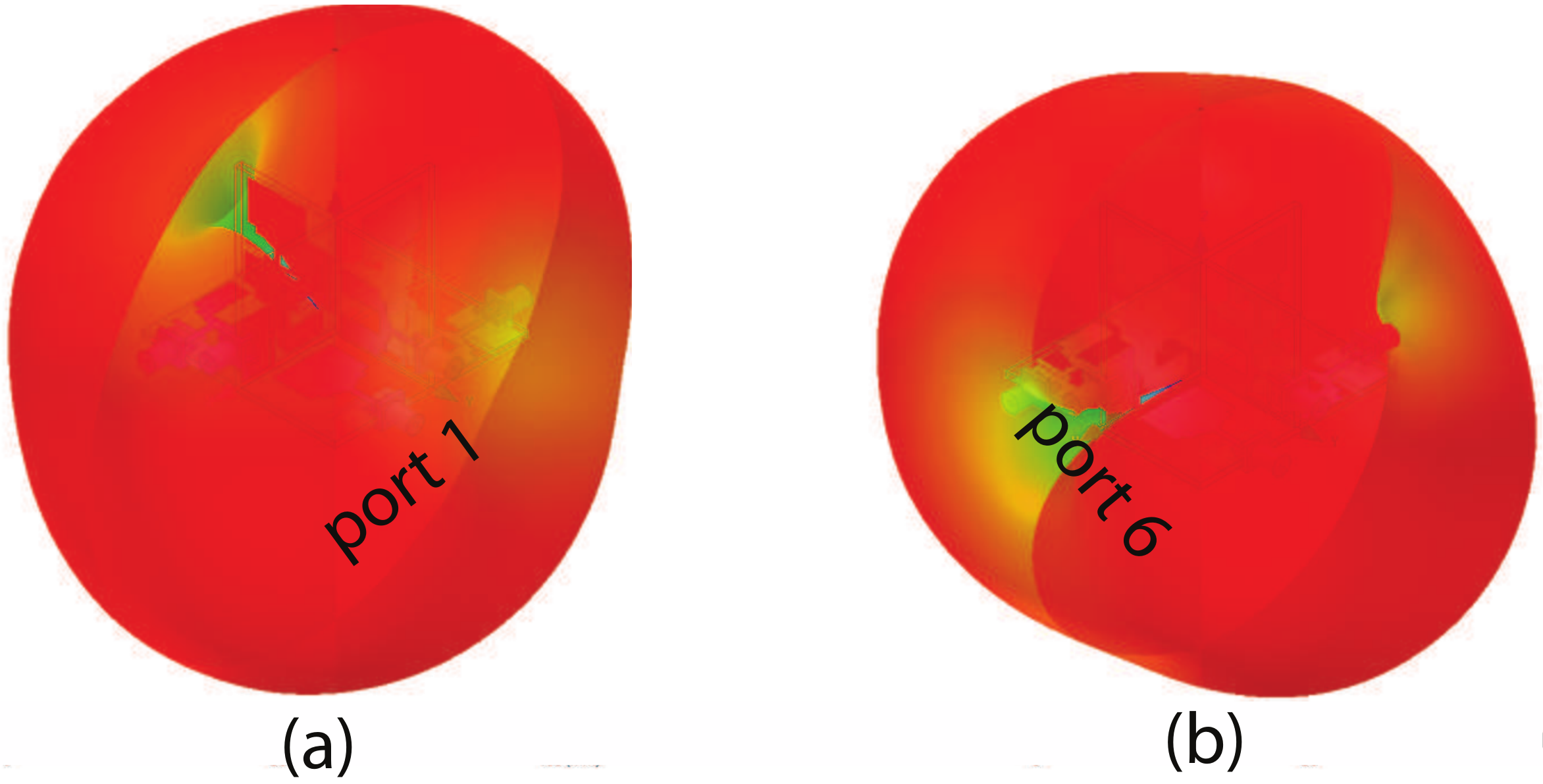}}
\caption{Simulated 3D radiation patterns showing pattern diversity in y-z plane for port1 and 6, (a) only port 1 is excited at 4 GHz, (b) only port 6 is excited at 4 GHz.}
\label{fig3d}
\end{figure}

\clearpage

\subsection{Radiation Patterns and Peak Gain}
The radiation patterns when feeding port 1 and 6 were measured at three different frequencies (3.5, 5.8 and 9.8 GHz) in  an anechoic chamber in their respective principle planes and compared with the simulated patterns. During the measurements, port 1 and 6 were excited one by one and all other ports were terminated with a 50$-\Omega$ matched load. The results for port 1 and 6 are plotted in Fig. 4 (A) and Fig. 4 (B), respectively. At the lower frequencies, the patterns are fairly dumbbell shaped in the y$-$z plane and omni directional in the x$-$z plane for port 1, as shown in Fig. 4 (A) (a)$-$(b). At 3.5 GHz, the agreement between simulated and measured results is fair. In the rejection band, the antenna radiates with very low intensity and gain in both planes, which is clearly visible in Fig. 4 (A) (c)$-$(d) at 5.8 GHz. At the higher frequencies, the pattern is slightly deviated from dumbbell shaped in the y$-$z plane and the discrepancies in both planes are slightly higher as compared to the lower frequencies due to more losses at higher frequencies. This is visible in Fig. 4 (A) (e)$-$(f), when the patterns are plotted at 9.8 GHz. As obvious from Fig. 1, monopole 1 is identical to monopole 3 and monopole 2 to monopole 4, so a mirror transformation at 180$^{\circ}$ in the radiation patterns in the respective planes is observed. Therefore, the patterns of port 2 are also a mirror transform in the perpendicular plane from port 1. Similarly, port 6 results (shown in Fig. 4 (B) (a)$-$(f)) exhibit almost the same behaviour in the z$-$y plane (omni directional) and x$-$y plane (dumbbell shape) as of port 1 in the x$-$z plane (omni directional) and y$-$z plane (dumbbell shape), respectively.

The simulated and measured gain at port 1 and port 6 are plotted in Fig. 5 (a). There are slight variations in the measured and simulated values and also from port 1 to port 6. Due to the identical monopoles, same values of gain are observed at other ports. The peak gain varies from 2.65 dBi to 5.8 dBi over the entire spectrum and it drops to $-$3.6 dB in the rejected band, showing that antennas are rejecting the band. The simulated and measured efficiency at port 1 and 6 is also plotted in Fig. 5 (b). It was observed that the efficiency was more than 86 $\%$ in the entire band except the rejected band.
\begin{figure}[!t]
\centerline{\includegraphics[width= 2.5 in]{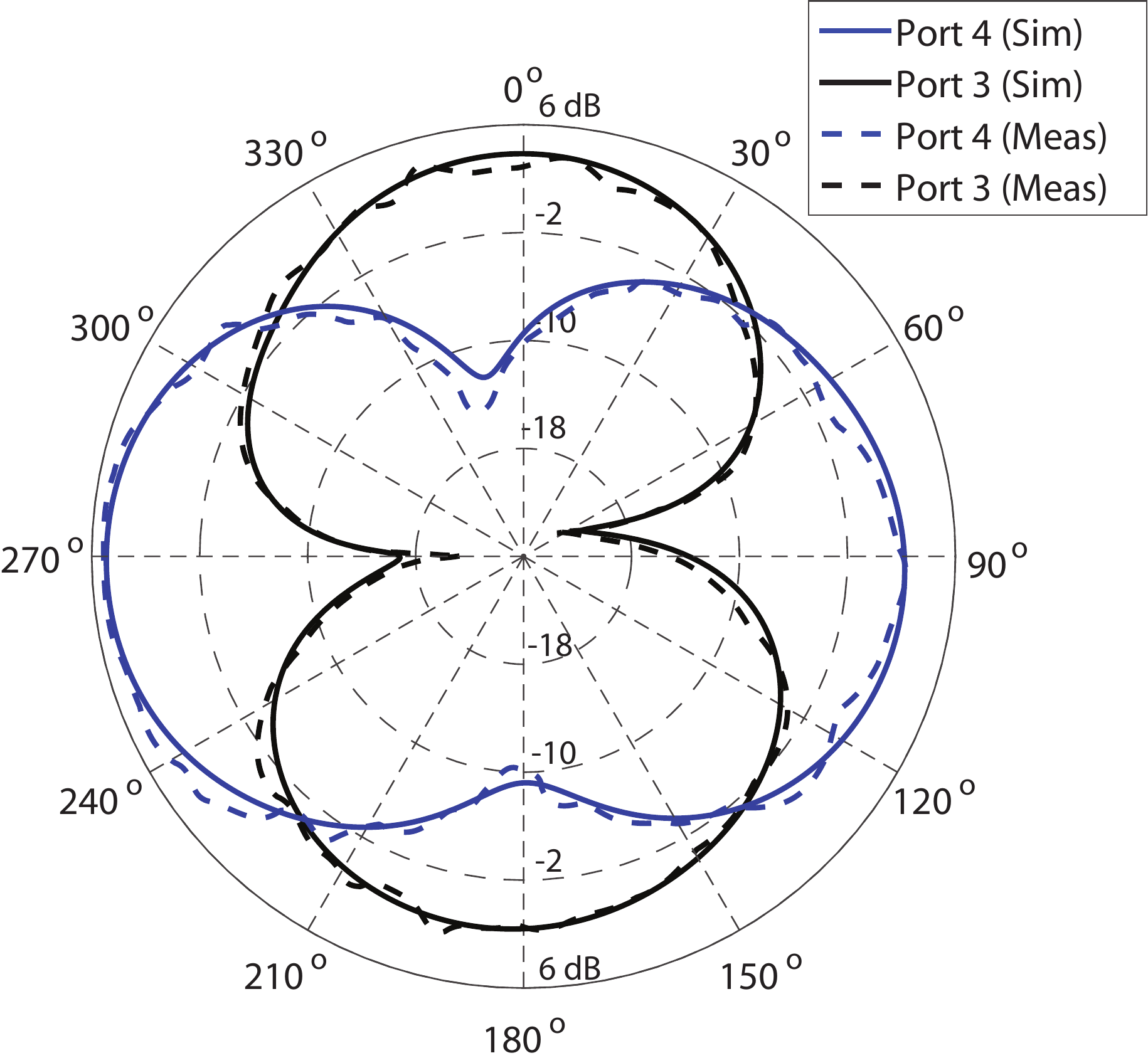}}
\caption{Simulated and measured radiation patterns of monopole pair 3$-$4 in the x-y plane at 4 GHz showing strong un-correlation at some angles, useful for diversity applications.}
\label{fig1e}
\end{figure}
\subsection{Diversity Analysis}
In MIMO systems, multipath effects can be mitigated, if different monopoles have patterns diversity in the respective plane. In the proposed design, monopole 1 has dumbbell shape (E$-$plane) radiation pattern in the y$-$z plane, while monopole 6 has omni directional (H$-$plane) radiation pattern in the y$-$z plane, as plotted in Fig 4 (A) and (B). This behavior can also be seen in Fig. 6 (a) and (b), where simulated 3D radiation patterns are plotted at 4 GHz for port 1 and 6, respectively. Similarly, to check the diversity amongst other monopoles, x$-$y plane radiation patterns of the monopole pair 3$-$4 are plotted in the Fig. 7 at 4 GHz. It is shown that monopole 3 has more radiation at 0$^{\circ}$ and 180$^{\circ}$ while monopole 4 has nulls in those directions, however monopole 4 has more radiation at 90$^{\circ}$ and 270$^{\circ}$, while monopole 3 has nulls in those directions. As a conclusion, these patterns are reasonably uncorrelated which is very well suited for the diversity applications.

\begin{figure}[!t]
\centerline{\includegraphics[width= \columnwidth]{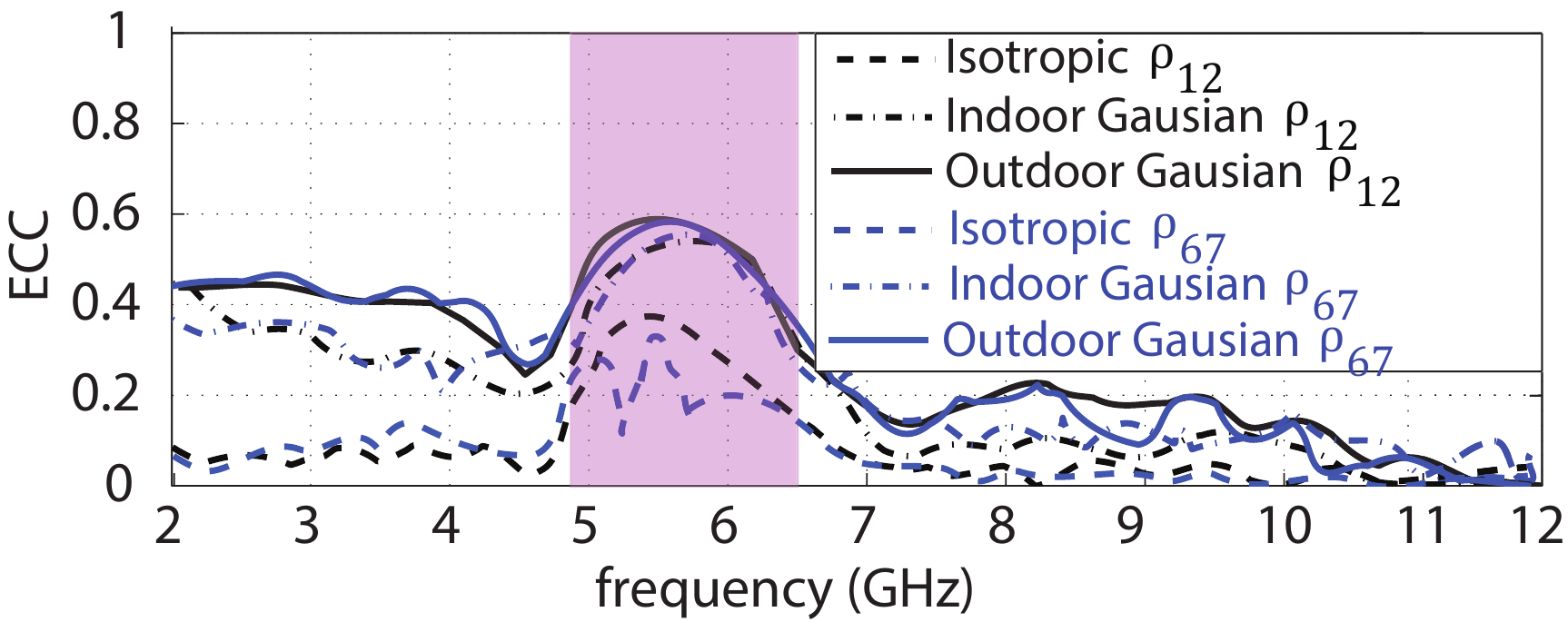}}
\caption{Computed envelope correlation coefficients from far-field radiation pattern for isotropic (uniform), indoor and outdoor environments. The XPR values used for indoor and outdoor environment are 5 dB and 1 dB, respectively. The ECC values are less than 0.45 in all cases.}
\label{fig1f}
\end{figure}

To further investigate, the diversity performance of the antenna is analysed by computing the envelope correlation coefficient (ECC) from far$-$field radiation patterns. For indoor and outdoor environments, the parameters defined in \cite{knudsen2002spherical} are used in (1) and ECC is numerically calculated. The computed ECC from the far$-$field radiation patterns is shown in Fig. 8. It can be observed from Fig. 8 that the computed values are less than 0.45 for all cases over the entire band. For uniform scattering environments, the values are less than 0.15.

\begin{equation}
\rho_{e}=\frac{\left|\int_{0}^{2\pi}\int_{0}^{\pi}\left(F_{\theta (1, 2),~\phi (1, 2)}~\right)d\Omega\right|^{2}}{\int_{0}^{2\pi}\int_{0}^{\pi}\left(F_{\theta (1, 2),~\phi (1, 2)}~\right)d\Omega\times\int_{0}^{2\pi}\int_{0}^{\pi}\left(F_{\theta2,~\phi2}~\right)d\Omega}.
\label{eq1}
\end{equation} Where,

\begin{equation}
F_{\theta (1, 2),~\varphi (1, 2)}= XPR.E_{\theta1}.E_{\theta2}^{\ast}.P_{\theta}+E_{\varphi1}.E_{\varphi2}^{\ast}.P_{\varphi}
\label{eq1a}
\end{equation} and
\begin{equation}
F_{\theta2,~\varphi2}= XPR.E_{\theta2}.E_{\theta2}^{\ast}.P_{\theta}+E_{\varphi2}.E_{\varphi2}^{\ast}.P_{\varphi}
\label{eq1b}
\end{equation}

\begin{figure}[h!]
\centerline{\includegraphics[width= \columnwidth]{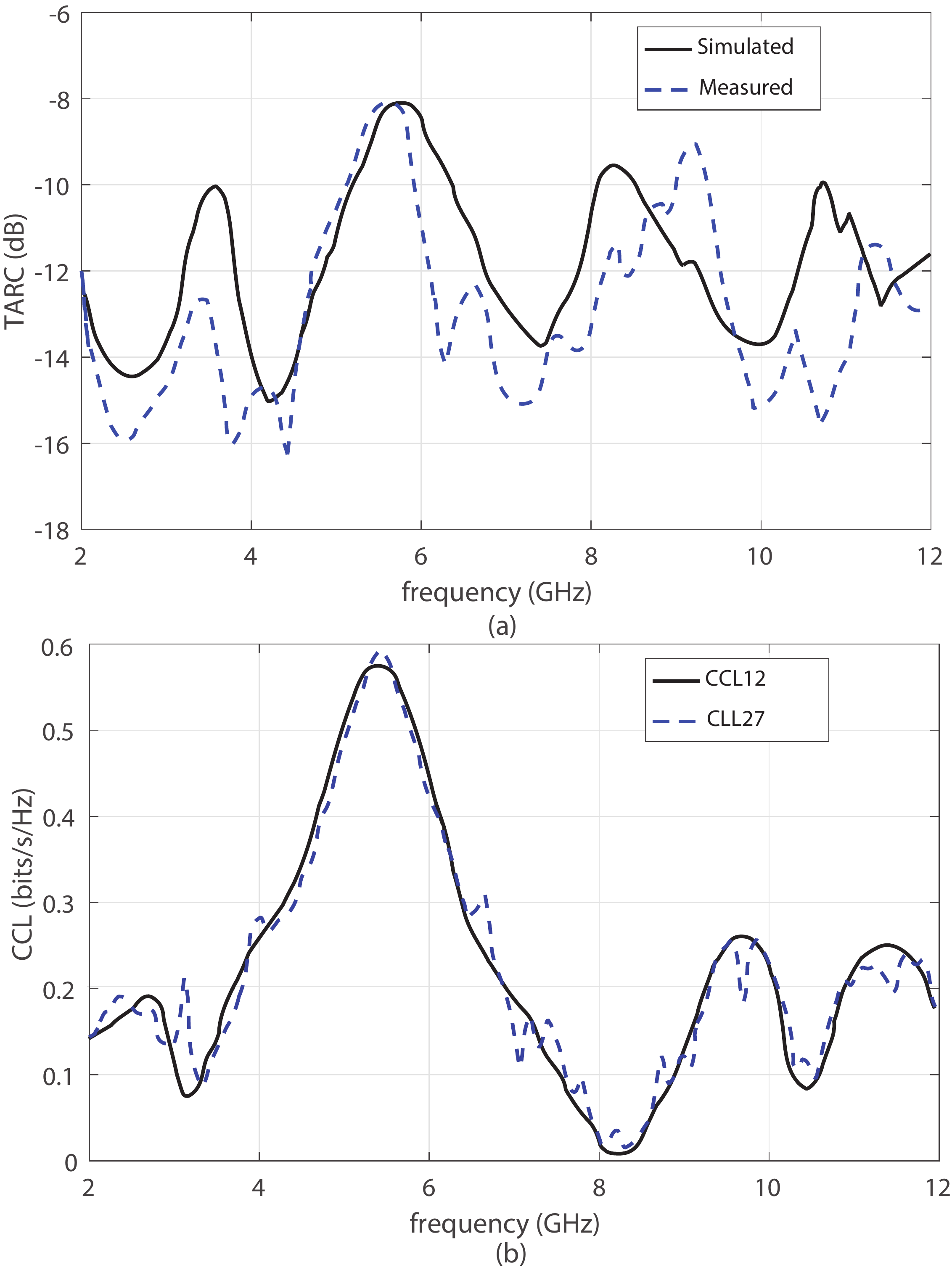}}
\caption{(a) Simulated and measured TARC computed from S-parameters. (b) CCL values between element 1, 2 and 2, 7.}
\label{figtarc}
\end{figure}

\begin{figure}[h!]
\centerline{\includegraphics[width= 2.5in]{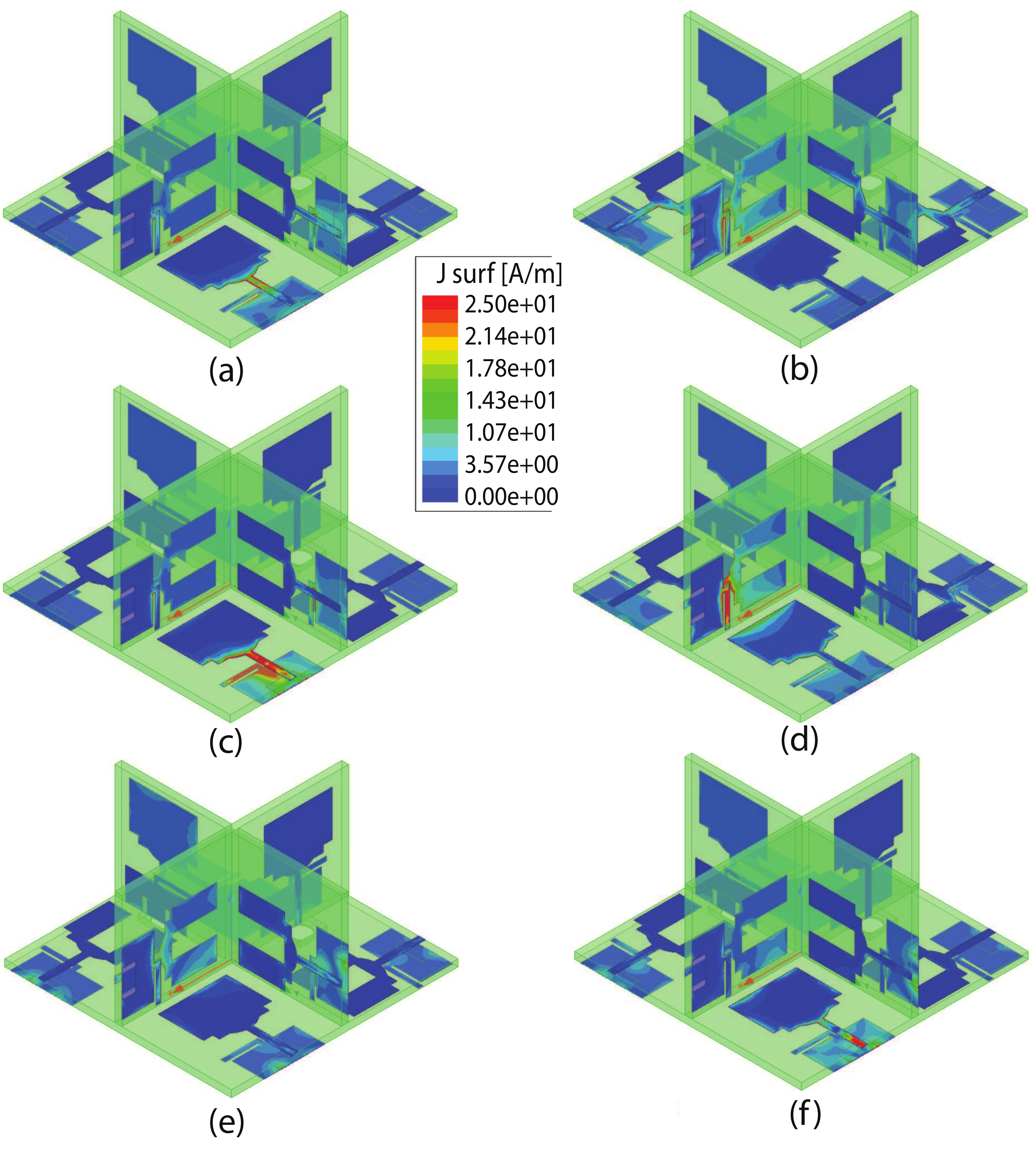}}
\caption{Surface current distribution at different frequencies. (a) exciting port 1 only at 3.5 GHz, and (b) exciting port 6 only at 3.5 GHz, (c) exciting port 1 only at 5.8 GHz, and (d) exciting port 6 only at 5.8 GHz, (e) exciting port 1 only at 9.8 GHz, and (f) exciting port 6 only at 9.8 GHz. The plot illustrates, that stub only draw current at 5.8 GHz to reject the band and does not affect others band much.}
\label{figcurrent}
\end{figure}

\begin{figure}[h!]
\centerline{\includegraphics[width= \columnwidth]{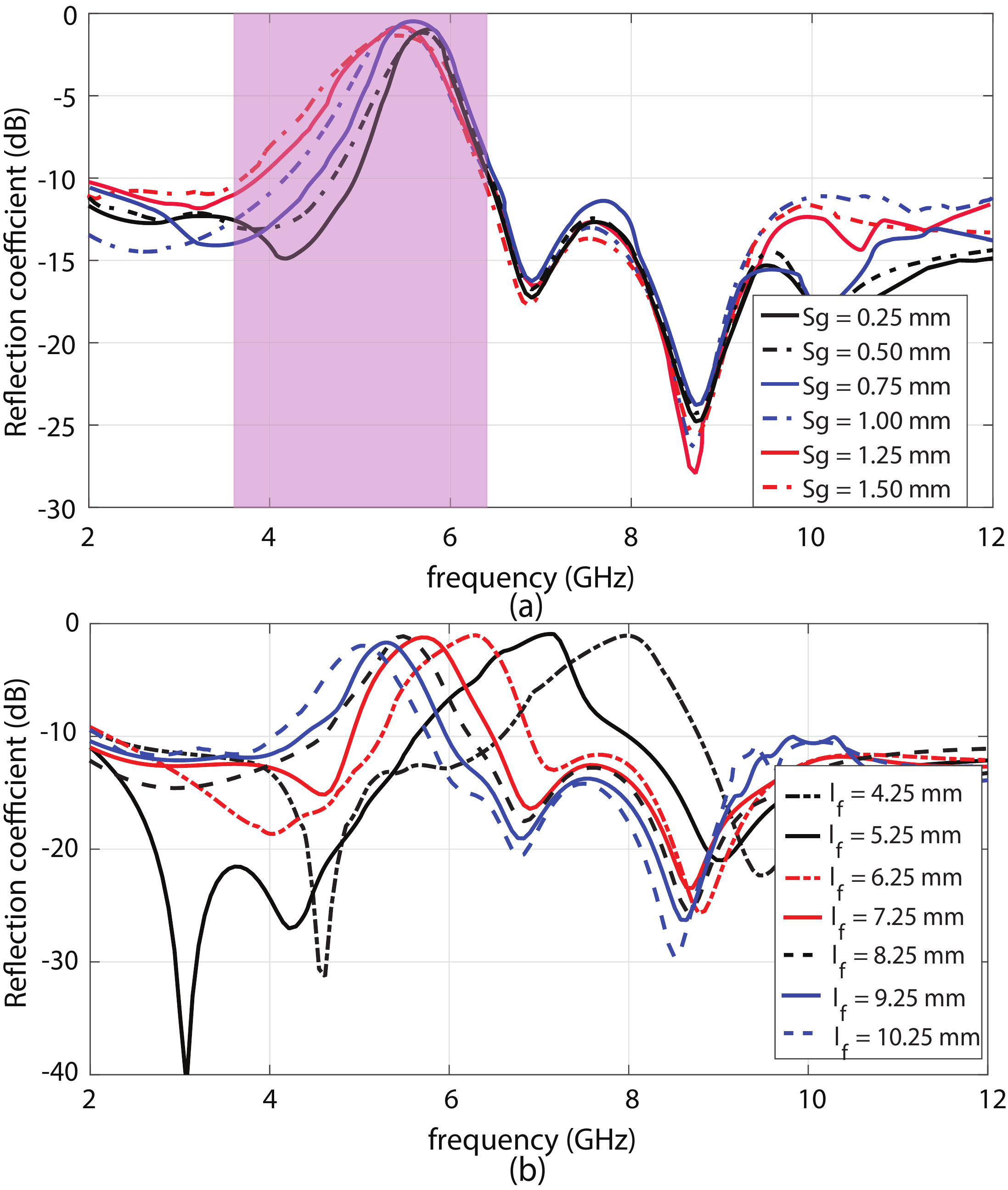}}
\caption{Effects of the stub on the reflection coefficient. (a) Effect of the gap variation between ground plane, the values are changed from 0.25 mm to 1.50 mm, the rejected band width can be controlled from 1 GHz to 2.6 GHz, (b) effects of the length variation of the stub, the values are changed from 4.25 mm to 10.25 mm, the rejected band shifted towards higher frequencies  and lower frequencies, when length is decreased and increased, respectively.}
\label{figbw}
\end{figure}

Next, Total Active Reflection Coefficient (TARC) and the Channel Capacity Loss (CCL) are computed using eq. (4) and (8) of \cite{wu2018quad}. For a desirable performance, TARC should be less than 0 dB while CCL should be no more than 0.5 bits/s/Hz. In the proposed design, the value of TARC was found to be less than -8 dB over the entire band, as shown in Fig. 9 (a) and CCL less than 0.3 bits/s/Hz except the reject band, as shown in Fig. 9 (b). Another important factor to evaluate the performance of MIMO antenna is Mean effective gain (MEG). The ratio obtained after calculation in different scenarios (such as isotropic, indoor and outdoor) was found close to 1.

Furthermore, to further show the effect of the stub on each monopole, the surface currents are plotted in Fig. 10 at different frequencies (3.5 GHz, 5.8 GHz, and 9.8 GHz) by exciting port 1 and 6. It can be noticed, when the surface currents are plotted at 3.5 GHz in Fig. 10 (a) and (b), while exciting port 1 and 6 individually, very little current is present on the stub, which does not impact the band at this frequency. Whereas, when surface currents are plotted at 5.8 GHz as shown in Fig. 10 (c) and (d), the current is present with high intensity on the stub, which behaves like a LC stop band filter and rejects the band. Similarly, when currents are plotted at 9.2 GHz for both ports excitations one by one as shown in Fig. 10 (e) and (f), the stub does not draw much current.

\clearpage

\section{Parametric Analysis On Stub Length and Gap Variation}
One major benefit of the proposed stub is that the bandwidth of the rejection can be optimized according to the requirement by changing the gap `$S_{g}$’ between the stub and ground plane. The results of gap variations are plotted in Fig. 11 (a). The gap variation changes the capacitance between stub and ground plane, resulting in the change of the rejection bandwidth. At the minimum gap 0.25 mm, the rejection bandwidth is 1 GHz (5.25 to 6.25 GHz), while at 1.5 mm gap, the bandwidth is around 2.6 GHz (3.7 to 6.3 GHz).  Also, the rejection band can be tuned to higher or lower frequencies by changing the length ‘lf’ of the stub as shown in Fig. 11 (b). The stub length is varied from 4.25 mm to 10.25 mm and shift in the rejection band is clearly visible in Fig. 11 (b).
\begin{table*}[t]
\caption{Performance Comparison With Previous Literature}
\scalebox{0.85}{
\begin{tabular}{|m{9em}|m{8em}|m{5em}|m{4em}|m{4.5em}|m{5.5em}|m{3em}|m{5em}|m{5em}|} \hline
 Published literature &  Total PCB size without height~/~No. of elements&  Bandwidth (GHz)&  Notch Band (GHz)& S$_{11}$ at notch (notch quality)&  Bandwidth control~/~Complexity& Mutual coupling (dB)&  Gain Var-(dBi)~/~Notch band gain (dBi)&  ECC using far-field patterns \\ \hline
 \cite{anitha2016compact} Anitha et al. &  45~\texttimes~45 mm$^{2}$~/~4 &  2.2$-$6.28 &No &N.A &N.A &$<-$14 &4~/~N.A   &	$<$ 0.25 (uniform)  \\ \hline
 \cite{lin2009ultra} Lin et al. &110~\texttimes~114 mm$^{2}$~/~4 & 2$-$6 & No & N.A & N.A & $<-$20 & 2.7~/~N.A & N.A \\ \hline
 \cite{khan2016compact} Asif et al. &50~\texttimes~39.8 mm$^{2}$ & 2.7$-$12 &4.8$-$6.2 &$-$2 dB &Not presented~/~complex design structure used to obtain notch & $<-17$ &4~/~$-$3.8 &N. A \\ \hline
 \cite{kiem2014design} Keim et al. &60~\texttimes~60 mm$^{2}$~/~4 & 2.73$-$10.68 & 5.36$-$6.04 & $-$1.8 dB &Not presented /complex EBG structure & $<-15$ & 5~/~$-$3.7 & N.A  \\ \hline
 \cite{wu2018quad} Wenjing et al. &60~\texttimes~60 mm$^{2}$~/~4 &3.0$-$16.2 &4.0$-$5.2 &$-$1.4 dB &Not presented /complex EBG structure with vias & $<-17.5$ & 6~/~$-$1 &$<$0.3 (uniform)  \\ \hline
 \cite{sharawi20135} Sharawi et al. &100~\texttimes~50 mm$^{2}$~/~8 &4.95$-$5.05 & No & N.A & N.A & $<-10.5$ & 0.8~/~N.A &N.A  \\ \hline
\cite{al2014eight} Al-Hadi et al. &110~\texttimes~55 mm$^{2}$~/~8 &3.4$-$3.6 & No & N.A & N.A & $<-10$ & N.A~/~N.A &N.A  \\ \hline
\cite{saleem2015eight} Saleem et al. & 60~\texttimes~93 mm$^{2}$~/~8 &3$-$10.6 & No & N.A & N.A & $<-15$ & N.A~/~N.A &N.A  \\ \hline
\cite{li2017eight} Li et al. & 68~\texttimes~136 mm$^{2}$~/~8 &2.5$-$2.6 & No & N.A & N.A & $<-15$ & 0.7~/~N.A &0.2 (uniform)  \\ \hline
\cite{sipal2017easily} Sipal et al. & 38~\texttimes~90 mm$^{2}$~/~8 &3$-$15 & No & N.A & N.A & $<-20$ & 4.5~/~N.A & N.A  \\ \hline
\cite{mathur20198} Rohit et al. & 85~\texttimes~85 mm$^{2}$~/~8 &3$-$10.6 & No & N.A & N.A & $<-15$ & 4.8~/~N.A & 0.2 (uniform)  \\ \hline
\cite{zhang2019ultra} Xugang et al. & 75~\texttimes~150 mm$^{2}$~/~8 &3$-$3.6 & No & N.A & N.A & $<-11$ & N.A~/~N.A & 0.1 (uniform)  \\ \hline
\cite{palaniswamy20163} Palaniswamy et al. & 70~\texttimes~70 mm$^{2}$~/~8 &2.9$-$12 & No & N.A & N.A & $<-16$ & N.A~/~N.A & 0.39  \\ \hline
\cite{alsath2018integrated} Alsath et al. & 90~\texttimes~90 mm$^{2}$~/~8 &3.1$-$12 & No & N.A & N.A & $<-17$ & 3.2~/~N.A & 0.16  \\ \hline
Proposed antenna &50~\texttimes~50 mm$^{2}$~/~8 &2$-$12 &4.85$-$6.35 &$-$1 dB &Yes~/~simple LC stub &$<-17$ &3.15~/~$-$3.6 &$<$0.15 (uniform) $<$0.45 (indoor) $<$0.45 (outdoor)  \\ \hline
\end{tabular}}
\end{table*}

In Table I, a comparison of the proposed antenna with several four element and eight element MIMO antenna designs is presented. The list is not complete but provides a reasonable understanding on the current work. Most of the eight element MIMO designs are only for narrow band operations, some designs do not provide band notch characteristics, some available designs provide band notch characteristics but for four element design. The proposed UWB-MIMO antenna has improved performance, when compared to all components of other antennas.

\clearpage

\section{Discussion}
It can be observed from the literature review and comparison table that though many MIMO antennas with orthogonal orientation exist with antennas placement horizontally and vertically, however the proposed design has additional benefits and novelty in terms of:
  \begin{itemize}
  \item Capability to reject the WLAN band in 3-D configuration of MIMO antenna with horizontal and vertical placement of antennas having orthogonal orientation.
  \item The planar configuration of the proposed design i.e. 50~\texttimes~50 mm$^{2}$ which is considered compact as compared to designs available in literature. For example, design presented in \cite{palaniswamy20163} has board size of 70~\texttimes~70 mm$^{2}$, whereas board size is 90~\texttimes~90 mm$^{2}$ in \cite{alsath2018integrated}. It is worth mentioning here that usually common ground plane antennas are preferred in MIMO/diversity applications, however in some application such as biomedical imaging using 5G, the common ground plane can be avoided to obtain the desired results but that depends only on the specific applications. Many disconnected ground plane MIMO/diversity antennas are commercially available \cite{admin_flatant_2018} for practical application and provide best possible outcome in certain environments where improved isolation is required such as LTE/WiMax mobile terminals \cite{li2014isolation}.
  \end{itemize}

The practical application of the proposed UWB-MIMO in a 3D configuration with disconnected ground plane can be for wireless communication in vehicular networks and imaging radars. The use of disconnected ground plane have been investigated for vehicular networks in \cite{palaniswamy20163}, \cite{kwon2016integrated}. In addition, disconnected ground plane UWB-MIMO antennas have been proposed for radar imaging applications \cite{bassi2013integrated}. Separate ground planes in a MIMO antenna array provide additional feature of compactness along with the suppression of near-field coupling for closely placed antenna elements. In addition, separate ground planes in a design removes the incorporation of complex decoupling structure that creates fabrication and hardware complexity. Therefore, other than 5G applications, the proposed design configuration can be used for communication in vehicular networks and radar imaging where multiple antennas are required in a limited confined space.

\clearpage

\section{Conclusion}
A compact eight port UWB-MIMO/diversity antenna having perpendicular placement of antenna element to utilize the height of CPE as compared to the available eight and ten port MIMO antennas is detailed in this report. Band rejection of all monopoles is accomplished by trapping current on an LC stub that is connected to the ground plane. Also, the proposed band-stop stub can be used to control the bandwidth of the rejected band by changing the gap between the stub and the ground plane, and can also shift the rejected frequency to upper band by decreasing the stub length and to lower band by increasing the stub length. Monopoles 1$-$4 are orthogonally placed on the same board to exploit the polarization diversity for high isolation. Monopoles 5$-$8 are placed orthogonally to the planar board (in between the monopoles 1-4), still exploiting the polarization diversity. The entire antenna measures only 50~\texttimes~50 mm$^{2}$ with all eight elements in planar board. The results of the fabricated prototype on FR4 laminate matches well with the simulated results of reflection coefficient, mutual coupling, peak gain and radiation patterns over the entire spectrum of 2 to 12 GHz. The simplicity, more elements in compact size, and good performance of the proposed design makes it a very strong candidate for small portable devices, vehicular network, vehicle to vehicle communication, and imaging radar. The work of this project adds to the previous contributions \cite{7696590,7928624,8231103,7305541,6512176,7696590,8888295,8292726,8530987,khan2016compact,khan2016compact1,khan2017ultra,omar2016uwb,shubair2015novel,shubair2015novel1,al2006direction,al2005direction,nwalozie2013simple,shubair2005robust,belhoul2003modelling,ibrahim2017compact,shubair2004robust,che2008propagation,el2016design,shubair1993closed,al2005computationally,al2003investigation,shubair2005performance,al2016millimeter,shubair2005convergence,khan2018triband,alharbi2018flexible,ibrahim2016reconfigurable,hakam2016novel,shubair1992simple,ghosal2018characteristic,khan2018novel,elsalamouny2015novel,khan2016pattern,shubair2005improved,8458184,7156466,7696430,7347957,8646600,7305294,8706527,7803806,7803808,9082193,alhajri2018accurate} reported in the literature in new applications and evolving technologies.

\section*{Acknowledgment}

This work was supported in part by the EU H2020 Marie Skłodowska-Curie Individual Fellowship (ViSionRF) under Grant 840854, by the MIUR (Italian Minister for Education) under initiative “Department of Excellence” (law 223/2016), and by COMSATS University, Islamabad under COMSATS Research Grant Program (Project~$\#$~16-63/CGRP/CUI/ISB/18/847).

\clearpage

\bibliographystyle{IEEEtran}
\bibliography{main}

\begin{thebibliography}{10}
\providecommand{\url}[1]{#1}
\csname url@samestyle\endcsname
\providecommand{\newblock}{\relax}
\providecommand{\bibinfo}[2]{#2}
\providecommand{\BIBentrySTDinterwordspacing}{\spaceskip=0pt\relax}
\providecommand{\BIBentryALTinterwordstretchfactor}{4}
\providecommand{\BIBentryALTinterwordspacing}{\spaceskip=\fontdimen2\font plus
\BIBentryALTinterwordstretchfactor\fontdimen3\font minus
  \fontdimen4\font\relax}
\providecommand{\BIBforeignlanguage}[2]{{%
\expandafter\ifx\csname l@#1\endcsname\relax
\typeout{** WARNING: IEEEtran.bst: No hyphenation pattern has been}%
\typeout{** loaded for the language `#1'. Using the pattern for}%
\typeout{** the default language instead.}%
\else
\language=\csname l@#1\endcsname
\fi
#2}}
\providecommand{\BIBdecl}{\relax}
\BIBdecl

\bibitem{kaiser2009overview}
T.~Kaiser, F.~Zheng, and E.~Dimitrov, ``An overview of ultra-wide-band systems
  with mimo,'' \emph{Proceedings of the IEEE}, vol.~97, no.~2, pp. 285--312,
  2009.

\bibitem{wallace2003experimental}
J.~W. Wallace, M.~A. Jensen, A.~L. Swindlehurst, and B.~D. Jeffs,
  ``Experimental characterization of the mimo wireless channel: Data
  acquisition and analysis,'' \emph{IEEE Transactions on Wireless
  Communications}, vol.~2, no.~2, pp. 335--343, 2003.

\bibitem{satam2018spanner}
V.~Satam, S.~Nema, and S.~S. Thakur, ``Spanner shape monopole mimo antenna with
  high gain for uwb applications,'' in \emph{Proceedings of International
  Conference on Wireless Communication}.\hskip 1em plus 0.5em minus 0.4em\relax
  Springer, 2018, pp. 129--138.

\bibitem{liu2013compact}
L.~Liu, S.~Cheung, and T.~Yuk, ``Compact mimo antenna for portable devices in
  uwb applications,'' \emph{IEEE Transactions on antennas and propagation},
  vol.~61, no.~8, pp. 4257--4264, 2013.

\bibitem{khan2015planar}
M.~S. Khan, A.-D. Capobianco, A.~Naqvi, B.~Ijaz, S.~Asif, and B.~D. Braaten,
  ``Planar, compact ultra-wideband polarisation diversity antenna array,''
  \emph{IET Microwaves, Antennas \& Propagation}, vol.~9, no.~15, pp.
  1761--1768, 2015.

\bibitem{anitha2016compact}
R.~Anitha, P.~Vinesh, K.~Prakash, P.~Mohanan, and K.~Vasudevan, ``A compact
  quad element slotted ground wideband antenna for mimo applications,''
  \emph{IEEE Transactions on Antennas and Propagation}, vol.~64, no.~10, pp.
  4550--4553, 2016.

\bibitem{lin2009ultra}
S.-Y. Lin and H.-R. Huang, ``Ultra-wideband mimo antenna with enhanced
  isolation,'' \emph{Microwave and Optical Technology Letters}, vol.~51, no.~2,
  pp. 570--573, 2009.

\bibitem{khan20154}
M.~S. Khan, A.-D. Capobianco, S.~Asif, A.~Iftikhar, and B.~D. Braaten, ``A 4
  element compact ultra-wideband mimo antenna array,'' in \emph{2015 IEEE
  International Symposium on Antennas and Propagation \& USNC/URSI National
  Radio Science Meeting}.\hskip 1em plus 0.5em minus 0.4em\relax IEEE, 2015,
  pp. 2305--2306.

\bibitem{khan2016compact}
M.~S. Khan, A.-D. Capobianco, S.~M. Asif, D.~E. Anagnostou, R.~M. Shubair, and
  B.~D. Braaten, ``A compact csrr-enabled uwb diversity antenna,'' \emph{IEEE
  Antennas and Wireless Propagation Letters}, vol.~16, pp. 808--812, 2016.

\bibitem{kiem2014design}
N.~K. Kiem, H.~N.~B. Phuong, and D.~N. Chien, ``Design of compact 4$\times$ 4
  uwb-mimo antenna with wlan band rejection,'' \emph{International Journal of
  Antennas and Propagation}, vol. 2014, 2014.

\bibitem{wu2018quad}
W.~Wu, B.~Yuan, and A.~Wu, ``A quad-element uwb-mimo antenna with band-notch
  and reduced mutual coupling based on ebg structures,'' \emph{International
  journal of Antennas and Propagation}, vol. 2018, 2018.

\bibitem{knudsen2002spherical}
M.~B. Knudsen and G.~F. Pedersen, ``Spherical outdoor to indoor power spectrum
  model at the mobile terminal,'' \emph{IEEE Journal on selected areas in
  communications}, vol.~20, no.~6, pp. 1156--1169, 2002.

\bibitem{sharawi20135}
M.~S. Sharawi, ``A 5-ghz 4/8-element mimo antenna system for ieee 802.11 ac
  devices,'' \emph{Microwave and Optical Technology Letters}, vol.~55, no.~7,
  pp. 1589--1594, 2013.

\bibitem{al2014eight}
A.~A. Al-Hadi, J.~Ilvonen, R.~Valkonen, and V.~Viikari, ``Eight-element antenna
  array for diversity and mimo mobile terminal in lte 3500 mhz band,''
  \emph{Microwave and Optical Technology Letters}, vol.~56, no.~6, pp.
  1323--1327, 2014.

\bibitem{saleem2015eight}
R.~Saleem, M.~Bilal, K.~Bajwa, and M.~Shafique, ``Eight-element uwb-mimo array
  with three distinct isolation mechanisms,'' \emph{Electronics Letters},
  vol.~51, no.~4, pp. 311--313, 2015.

\bibitem{li2017eight}
M.-Y. Li, Z.-Q. Xu, Y.-L. Ban, Z.-F. Yu \emph{et~al.}, ``Eight-port
  orthogonally dual-polarised mimo antennas using loop structures for 5g
  smartphone,'' \emph{IET Microwaves, Antennas \& Propagation}, vol.~11,
  no.~12, pp. 1810--1816, 2017.

\bibitem{sipal2017easily}
D.~Sipal, M.~P. Abegaonkar, and S.~K. Koul, ``Easily extendable compact planar
  uwb mimo antenna array,'' \emph{IEEE Antennas and Wireless Propagation
  Letters}, vol.~16, pp. 2328--2331, 2017.

\bibitem{mathur20198}
R.~Mathur and S.~Dwari, ``8-port multibeam planar uwb-mimo antenna with pattern
  and polarisation diversity,'' \emph{IET Microwaves, Antennas \& Propagation},
  vol.~13, no.~13, pp. 2297--2302, 2019.

\bibitem{zhang2019ultra}
X.~Zhang, Y.~Li, W.~Wang, and W.~Shen, ``Ultra-wideband 8-port mimo antenna
  array for 5g metal-frame smartphones,'' \emph{IEEE access}, vol.~7, pp.
  72\,273--72\,282, 2019.

\bibitem{anagnostou2007dual}
D.~E. Anagnostou, S.~Nikolaou, H.~Kim, B.~Kim, M.~Tentzeris, and
  J.~Papapolymerou, ``Dual band-notched ultra-wideband antenna for 802.11 a
  wlan environments,'' in \emph{2007 IEEE Antennas and Propagation Society
  International Symposium}.\hskip 1em plus 0.5em minus 0.4em\relax IEEE, 2007,
  pp. 4621--4624.

\bibitem{gheethan2012dual}
A.~A. Gheethan and D.~E. Anagnostou, ``Dual band-reject uwb antenna with sharp
  rejection of narrow and closely-spaced bands,'' \emph{IEEE transactions on
  antennas and propagation}, vol.~60, no.~4, pp. 2071--2076, 2012.

\bibitem{anagnostou2013reconfigurable}
D.~E. Anagnostou, M.~T. Chryssomallis, B.~D. Braaten, J.~L. Ebel, and
  N.~Sep{\'u}lveda, ``Reconfigurable uwb antenna with rf-mems for on-demand
  wlan rejection,'' \emph{IEEE Transactions on Antennas and Propagation},
  vol.~62, no.~2, pp. 602--608, 2013.

\bibitem{ban20164g}
Y.-L. Ban, C.~Li, G.~Wu, K.-L. Wong \emph{et~al.}, ``4g/5g multiple antennas
  for future multi-mode smartphone applications,'' \emph{IEEE access}, vol.~4,
  pp. 2981--2988, 2016.

\bibitem{li2016eight}
M.-Y. Li, Y.-L. Ban, Z.-Q. Xu, G.~Wu, K.~Kang, Z.-F. Yu \emph{et~al.},
  ``Eight-port orthogonally dual-polarized antenna array for 5g smartphone
  applications,'' \emph{IEEE Transactions on Antennas and Propagation},
  vol.~64, no.~9, pp. 3820--3830, 2016.

\bibitem{guo2018side}
J.~Guo, L.~Cui, C.~Li, and B.~Sun, ``Side-edge frame printed eight-port
  dual-band antenna array for 5g smartphone applications,'' \emph{IEEE
  Transactions on Antennas and Propagation}, vol.~66, no.~12, pp. 7412--7417,
  2018.

\bibitem{li201712}
Y.~Li, Y.~Luo, G.~Yang \emph{et~al.}, ``12-port 5g massive mimo antenna array
  in sub-6ghz mobile handset for lte bands 42/43/46 applications,'' \emph{IEEE
  access}, vol.~6, pp. 344--354, 2017.

\bibitem{li2018multiband}
------, ``Multiband 10-antenna array for sub-6 ghz mimo applications in 5-g
  smartphones,'' \emph{IEEE Access}, vol.~6, pp. 28\,041--28\,053, 2018.

\bibitem{khan2020ultra}
M.~S. Khan, A.~Iftikhar, R.~M. Shubair, A.-D. Capobianco, S.~M. Asif, B.~D.
  Braaten, and D.~E. Anagnostou, ``Ultra-compact reconfigurable band reject uwb
  mimo antenna with four radiators,'' \emph{Electronics}, vol.~9, no.~4, p.
  584, 2020.

\bibitem{khan2018compact}
M.~Khan, F.~Rigobello, B.~Ijaz, E.~Autizi, A.~Capobianco, R.~Shubair, and
  S.~Khan, ``Compact 3-d eight elements uwb-mimo array,'' \emph{Microwave and
  Optical Technology Letters}, vol.~60, no.~8, pp. 1967--1971, 2018.

\bibitem{7696590}
M.~S. {Khan}, A.~. {Capobianco}, S.~M. {Asif}, A.~{Iftikhar}, B.~D. {Braaten},
  and R.~M. {Shubair}, ``A properties comparison between copper and
  graphene-based uwb mimo planar antennas,'' in \emph{2016 IEEE International
  Symposium on Antennas and Propagation (APSURSI)}, 2016, pp. 1767--1768.

\bibitem{7928624}
A.~S. {Fazal}, U.~{Nasir}, B.~{Ijaz}, K.~S. {Alimgeer}, M.~F. {Shafique}, R.~M.
  {Shubair}, and M.~S. {Khan}, ``A compact uwb cpw-fed antenna with inverted
  l-shaped slot for wlan band notched characteristics,'' in \emph{2017 11th
  European Conference on Antennas and Propagation (EUCAP)}, 2017, pp. 981--984.

\bibitem{8231103}
G.~{Mansutti}, F.~{Rigobello}, M.~S. {Khan}, and A.~{Capobianco}, ``Pattern
  recovery of a beam-tilting phased array antenna on a doubly wedge-deformed
  surface,'' in \emph{2017 47th European Microwave Conference (EuMC)}, 2017,
  pp. 1353--1356.

\bibitem{7305541}
M.~S. {Khan}, A.~{Capobianco}, S.~{Asif}, A.~{Iftikhar}, and B.~D. {Braaten},
  ``A 4 element compact ultra-wideband mimo antenna array,'' in \emph{2015 IEEE
  International Symposium on Antennas and Propagation USNC/URSI National Radio
  Science Meeting}, 2015, pp. 2305--2306.

\bibitem{6512176}
M.~S. {Khan}, M.~F. {Shafique}, A.~D. {Capobianco}, E.~{Autizi}, and
  I.~{Shoaib}, ``Compact uwb-mimo antenna array with a novel decoupling
  structure,'' in \emph{Proceedings of 2013 10th International Bhurban
  Conference on Applied Sciences Technology (IBCAST)}, 2013, pp. 347--350.

\bibitem{8888295}
U.~{Farooq}, A.~{Iftikhar}, M.~J. {Mughal}, M.~{Farhan Shafique}, M.~S. {Khan},
  and R.~M. {Shubair}, ``A compact metasurface based cross polarization
  converter for x band applications,'' in \emph{2019 IEEE International
  Symposium on Antennas and Propagation and USNC-URSI Radio Science Meeting},
  2019, pp. 723--724.

\bibitem{8292726}
T.~{Asghar}, B.~{Ijaz}, K.~S. {Alimgeer}, M.~S. {Khan}, and R.~{Shubair}, ``A
  compact uwb mimo antenna with inverted u-shaped slot for wlan rejection,'' in
  \emph{2017 IEEE 28th Annual International Symposium on Personal, Indoor, and
  Mobile Radio Communications (PIMRC)}, 2017, pp. 1--4.

\bibitem{8530987}
B.~{Ijaz}, A.~{Iftikhar}, K.~S. {Alimgeer}, M.~S. {Khan}, and R.~{Shubair}, ``A
  frequency reconfigurable dual-band monopole antenna for wireless
  applications,'' in \emph{2018 International Symposium on Networks, Computers
  and Communications (ISNCC)}, 2018, pp. 1--5.

\bibitem{khan2016compact1}
M.~S. Khan, A.-D. Capobianco, A.~Iftikhar, S.~Asif, and B.~D. Braaten, ``A
  compact dual polarized ultrawideband multiple-input-multiple-output
  antenna,'' \emph{Microwave and Optical Technology Letters}, vol.~58, no.~1,
  pp. 163--166, 2016.

\bibitem{khan2017ultra}
M.~S. Khan, A.-D. Capobianco, A.~Iftikhar, R.~M. Shubair, D.~E. Anagnostou, and
  B.~D. Braaten, ``Ultra-compact dual-polarised uwb mimo antenna with meandered
  feeding lines,'' \emph{IET Microwaves, Antennas \& Propagation}, vol.~11,
  no.~7, pp. 997--1002, 2017.

\bibitem{omar2016uwb}
A.~Omar and R.~Shubair, ``Uwb coplanar waveguide-fed-coplanar strips spiral
  antenna,'' in \emph{2016 10th European Conference on Antennas and Propagation
  (EuCAP)}.\hskip 1em plus 0.5em minus 0.4em\relax IEEE, 2016, pp. 1--2.

\bibitem{shubair2015novel}
R.~M. Shubair, A.~Salah, and A.~K. Abbas, ``Novel implantable miniaturized
  circular microstrip antenna for biomedical telemetry,'' in \emph{2015 IEEE
  International Symposium on Antennas and Propagation \& USNC/URSI National
  Radio Science Meeting}.\hskip 1em plus 0.5em minus 0.4em\relax IEEE, 2015,
  pp. 947--948.

\bibitem{shubair2015novel1}
R.~M. Shubair, A.~M. AlShamsi, K.~Khalaf, and A.~Kiourti, ``Novel miniature
  wearable microstrip antennas for ism-band biomedical telemetry,'' in
  \emph{2015 Loughborough Antennas \& Propagation Conference (LAPC)}.\hskip 1em
  plus 0.5em minus 0.4em\relax IEEE, 2015, pp. 1--4.

\bibitem{al2006direction}
E.~M. Al-Ardi, R.~M. Shubair, and M.~E. Al-Mualla, ``Direction of arrival
  estimation in a multipath environment: An overview and a new contribution,''
  \emph{Applied Computational Electromagnetics Society Journal}, vol.~21,
  no.~3, p. 226, 2006.

\bibitem{al2005direction}
M.~Al-Nuaimi, R.~Shubair, and K.~Al-Midfa, ``Direction of arrival estimation in
  wireless mobile communications using minimum variance distortionless
  response,'' in \emph{The Second International Conference on Innovations in
  Information Technology (IIT’05)}, 2005, pp. 1--5.

\bibitem{nwalozie2013simple}
G.~Nwalozie, V.~Okorogu, S.~Maduadichie, and A.~Adenola, ``A simple comparative
  evaluation of adaptive beam forming algorithms,'' \emph{International Journal
  of Engineering and Innovative Technology (IJEIT)}, vol.~2, no.~7, 2013.

\bibitem{shubair2005robust}
R.~Shubair, ``Robust adaptive beamforming using lms algorithm with smi
  initialization,'' in \emph{2005 IEEE Antennas and Propagation Society
  International Symposium}, vol.~4.\hskip 1em plus 0.5em minus 0.4em\relax
  IEEE, 2005, pp. 2--5.

\bibitem{belhoul2003modelling}
F.~A. Belhoul, R.~M. Shubair, and M.~Ai-Mualla, ``Modelling and performance
  analysis of doa estimation in adaptive signal processing arrays,'' in
  \emph{10th IEEE International Conference on Electronics, Circuits and
  Systems, 2003. ICECS 2003. Proceedings of the 2003}, vol.~1.\hskip 1em plus
  0.5em minus 0.4em\relax IEEE, 2003, pp. 340--343.

\bibitem{ibrahim2017compact}
A.~A. Ibrahim, J.~Machac, and R.~M. Shubair, ``Compact uwb mimo antenna with
  pattern diversity and band rejection characteristics,'' \emph{Microwave and
  Optical Technology Letters}, vol.~59, no.~6, pp. 1460--1464, 2017.

\bibitem{shubair2004robust}
R.~Shubair and A.~Al-Merri, ``Robust algorithms for direction finding and
  adaptive beamforming: performance and optimization,'' in \emph{The 2004 47th
  Midwest Symposium on Circuits and Systems, 2004. MWSCAS'04.}, vol.~2.\hskip
  1em plus 0.5em minus 0.4em\relax IEEE, 2004, pp. II--II.

\bibitem{che2008propagation}
W.~Che, C.~Li, P.~Russer, and Y.~Chow, ``Propagation and band broadening effect
  of planar integrated ridged waveguide in multilayer dielectric substrates,''
  in \emph{2008 IEEE MTT-S International Microwave Symposium Digest}.\hskip 1em
  plus 0.5em minus 0.4em\relax IEEE, 2008, pp. 217--220.

\bibitem{el2016design}
M.~El~Shorbagy, R.~M. Shubair, M.~I. AlHajri, and N.~K. Mallat, ``On the design
  of millimetre-wave antennas for 5g,'' in \emph{2016 16th Mediterranean
  Microwave Symposium (MMS)}.\hskip 1em plus 0.5em minus 0.4em\relax IEEE,
  2016, pp. 1--4.

\bibitem{shubair1993closed}
R.~Shubair and Y.~Chow, ``A closed-form solution of vertical dipole antennas
  above a dielectric half-space,'' \emph{IEEE transactions on antennas and
  propagation}, vol.~41, no.~12, pp. 1737--1741, 1993.

\bibitem{al2005computationally}
E.~M. Al-Ardi, R.~M. Shubair, and M.~E. Al-Mualla, ``Computationally efficient
  doa estimation in a multipath environment using covariance differencing and
  iterative spatial smoothing,'' in \emph{2005 IEEE International Symposium on
  Circuits and Systems}.\hskip 1em plus 0.5em minus 0.4em\relax IEEE, 2005, pp.
  3805--3808.

\bibitem{al2003investigation}
E.~Al-Ardi, R.~Shubair, and M.~Al-Mualla, ``Investigation of high-resolution
  doa estimation algorithms for optimal performance of smart antenna systems,''
  2003.

\bibitem{shubair2005performance}
R.~Shubair and W.~Jessmi, ``Performance analysis of smi adaptive beamforming
  arrays for smart antenna systems,'' in \emph{2005 IEEE Antennas and
  Propagation Society International Symposium}, vol.~1.\hskip 1em plus 0.5em
  minus 0.4em\relax IEEE, 2005, pp. 311--314.

\bibitem{al2016millimeter}
F.~Al-Ogaili and R.~M. Shubair, ``Millimeter-wave mobile communications for 5g:
  Challenges and opportunities,'' in \emph{2016 IEEE International Symposium on
  Antennas and Propagation (APSURSI)}.\hskip 1em plus 0.5em minus 0.4em\relax
  IEEE, 2016, pp. 1003--1004.

\bibitem{shubair2005convergence}
R.~Shubair and A.~Merri, ``Convergence of adaptive beamforming algorithms for
  wireless communications,'' in \emph{Proc. IEEE and IFIP International
  Conference on Wireless and Optical Communications Networks}, 2005, pp. 6--8.

\bibitem{khan2018triband}
O.~M. Khan, R.~M. Shubair, Q.~U. Islam, and I.~Rashid, ``Triband metamaterial
  embedded implantable antenna for biotelemetry applications,'' in \emph{2018
  IEEE International Symposium on Antennas and Propagation \& USNC/URSI
  National Radio Science Meeting}.\hskip 1em plus 0.5em minus 0.4em\relax IEEE,
  2018, pp. 213--214.

\bibitem{alharbi2018flexible}
S.~Alharbi, R.~M. Shubair, and A.~Kiourti, ``Flexible antennas for wearable
  applications: Recent advances and design challenges,'' 2018.

\bibitem{ibrahim2016reconfigurable}
A.~A. Ibrahim and R.~M. Shubair, ``Reconfigurable band-notched uwb antenna for
  cognitive radio applications,'' in \emph{2016 16th Mediterranean Microwave
  Symposium (MMS)}.\hskip 1em plus 0.5em minus 0.4em\relax IEEE, 2016, pp.
  1--4.

\bibitem{hakam2016novel}
A.~Hakam, M.~Hussein, M.~Ouda, R.~Shubair, and E.~Serria, ``Novel circular
  antenna with elliptical rings for ultra-wide-band,'' in \emph{2016 10th
  European Conference on Antennas and Propagation (EuCAP)}.\hskip 1em plus
  0.5em minus 0.4em\relax IEEE, 2016, pp. 1--4.

\bibitem{shubair1992simple}
R.~Shubair and Y.~Chow, ``A simple and accurate approach to model the coupling
  of vertical and horizontal dipoles in layered media,'' in \emph{IEEE Antennas
  and Propagation Society International Symposium 1992 Digest}.\hskip 1em plus
  0.5em minus 0.4em\relax IEEE, 1992, pp. 2309--2312.

\bibitem{ghosal2018characteristic}
S.~Ghosal, A.~De, A.~Chakrabarty, and R.~M. Shubair, ``Characteristic mode
  analysis of slot loading in microstrip patch antenna,'' in \emph{2018 IEEE
  International Symposium on Antennas and Propagation \& USNC/URSI National
  Radio Science Meeting}.\hskip 1em plus 0.5em minus 0.4em\relax IEEE, 2018,
  pp. 1523--1524.

\bibitem{khan2018novel}
O.~M. Khan, Q.~U. Islam, R.~M. Shubair, and A.~Kiourti, ``Novel multiband
  flamenco fractal antenna for wearable wban off-body communication
  applications,'' in \emph{2018 International Applied Computational
  Electromagnetics Society Symposium (ACES)}.\hskip 1em plus 0.5em minus
  0.4em\relax IEEE, 2018, pp. 1--2.

\bibitem{elsalamouny2015novel}
M.~Y. ElSalamouny and R.~M. Shubair, ``Novel design of compact low-profile
  multi-band microstrip antennas for medical applications,'' in \emph{2015
  loughborough antennas \& propagation conference (LAPC)}.\hskip 1em plus 0.5em
  minus 0.4em\relax IEEE, 2015, pp. 1--4.

\bibitem{khan2016pattern}
M.~S. Khan, A.-D. Capobianco, S.~M. Asif, A.~Iftikhar, B.~D. Braaten, and R.~M.
  Shubair, ``A pattern reconfigurable printed patch antenna,'' in \emph{2016
  IEEE International Symposium on Antennas and Propagation (APSURSI)}.\hskip
  1em plus 0.5em minus 0.4em\relax IEEE, 2016, pp. 2149--2150.

\bibitem{shubair2005improved}
R.~Shubair, A.~Merri, and W.~Jessmi, ``Improved adaptive beamforming using a
  hybrid lms/smi approach,'' in \emph{Second IFIP International Conference on
  Wireless and Optical Communications Networks (WOCN)}.\hskip 1em plus 0.5em
  minus 0.4em\relax IEEE, 2005, pp. 603--606.

\bibitem{8458184}
M.~I. {AlHajri}, N.~T. {Ali}, and R.~M. {Shubair}, ``Classification of indoor
  environments for iot applications: A machine learning approach,'' \emph{IEEE
  Antennas and Wireless Propagation Letters}, vol.~17, no.~12, pp. 2164--2168,
  2018.

\bibitem{7156466}
M.~I. {AlHajri}, A.~{Goian}, M.~{Darweesh}, R.~{AlMemari}, R.~M. {Shubair},
  L.~{Weruaga}, and A.~R. {Kulaib}, ``Hybrid rss-doa technique for enhanced wsn
  localization in a correlated environment,'' in \emph{2015 International
  Conference on Information and Communication Technology Research (ICTRC)},
  2015, pp. 238--241.

\bibitem{7696430}
M.~I. {AlHajri}, N.~{Alsindi}, N.~T. {Ali}, and R.~M. {Shubair},
  ``Classification of indoor environments based on spatial correlation of rf
  channel fingerprints,'' in \emph{2016 IEEE International Symposium on
  Antennas and Propagation (APSURSI)}, 2016, pp. 1447--1448.

\bibitem{7347957}
A.~{Goian}, M.~I. {AlHajri}, R.~M. {Shubair}, L.~{Weruaga}, A.~R. {Kulaib},
  R.~{AlMemari}, and M.~{Darweesh}, ``Fast detection of coherent signals using
  pre-conditioned root-music based on toeplitz matrix reconstruction,'' in
  \emph{2015 IEEE 11th International Conference on Wireless and Mobile
  Computing, Networking and Communications (WiMob)}, 2015, pp. 168--174.

\bibitem{8646600}
M.~I. {AlHajri}, N.~T. {Ali}, and R.~M. {Shubair}, ``A machine learning
  approach for the classification of indoor environments using rf signatures,''
  in \emph{2018 IEEE Global Conference on Signal and Information Processing
  (GlobalSIP)}, 2018, pp. 1060--1062.

\bibitem{7305294}
M.~I. {AlHajri}, R.~M. {Shubair}, L.~{Weruaga}, A.~R. {Kulaib}, A.~{Goian},
  M.~{Darweesh}, and R.~{AlMemari}, ``Hybrid method for enhanced detection of
  coherent signals using circular antenna arrays,'' in \emph{2015 IEEE
  International Symposium on Antennas and Propagation USNC/URSI National Radio
  Science Meeting}, 2015, pp. 1810--1811.

\bibitem{8706527}
M.~I. {AlHajri}, N.~T. {Ali}, and R.~M. {Shubair}, ``Indoor localization for
  iot using adaptive feature selection: A cascaded machine learning approach,''
  \emph{IEEE Antennas and Wireless Propagation Letters}, vol.~18, no.~11, pp.
  2306--2310, 2019.

\bibitem{7803806}
R.~M. {Shubair}, A.~S. {Goian}, M.~I. {AlHajri}, and A.~R. {Kulaib}, ``A new
  technique for uca-based doa estimation of coherent signals,'' in \emph{2016
  16th Mediterranean Microwave Symposium (MMS)}, 2016, pp. 1--3.

\bibitem{7803808}
R.~{Karli}, H.~{Ammor}, R.~M. {Shubair}, M.~I. {AlHajri}, R.~{Alkurd}, and
  A.~{Hakam}, ``Miniature planar ultra-wide-band microstrip antenna for breast
  cancer detection,'' in \emph{2016 16th Mediterranean Microwave Symposium
  (MMS)}, 2016, pp. 1--4.

\bibitem{9082193}
Z.~{Chen}, M.~I. {AlHajri}, M.~{Wu}, N.~T. {Ali}, and R.~M. {Shubair}, ``A
  novel real-time deep learning approach for indoor localization based on rf
  environment identification,'' \emph{IEEE Sensors Letters}, vol.~4, no.~6, pp.
  1--4, 2020.

\bibitem{alhajri2018accurate}
M.~AlHajri, A.~Goian, M.~Darweesh, R.~AlMemari, R.~Shubair, L.~Weruaga, and
  A.~AlTunaiji, ``Accurate and robust localization techniques for wireless
  sensor networks,'' \emph{arXiv preprint arXiv:1806.05765}, 2018.

\bibitem{palaniswamy20163}
S.~K. Palaniswamy, Y.~P. Selvam, M.~G.~N. Alsath, M.~Kanagasabai, S.~Kingsly,
  and S.~Subbaraj, ``3-d eight-port ultrawideband antenna array for diversity
  applications,'' \emph{IEEE Antennas and Wireless Propagation Letters},
  vol.~16, pp. 569--572, 2016.

\bibitem{alsath2018integrated}
M.~G.~N. Alsath, H.~Arun, Y.~P. Selvam, M.~Kanagasabai, S.~Kingsly,
  S.~Subbaraj, R.~Sivasamy, S.~K. Palaniswamy, and R.~Natarajan, ``An
  integrated tri-band/uwb polarization diversity antenna for vehicular
  networks,'' \emph{IEEE Transactions on Vehicular Technology}, vol.~67, no.~7,
  pp. 5613--5620, 2018.

\bibitem{admin_flatant_2018}
\BIBentryALTinterwordspacing
admin, ``\BIBforeignlanguage{en-US}{Flatant 8×8},'' Dec. 2018, library
  Catalog: compex.com.sg. [Online]. Available:
  \url{https://compex.com.sg/shop/antenna/flatant-8x8/}
\BIBentrySTDinterwordspacing

\bibitem{li2014isolation}
G.~Li, H.~Zhai, Z.~Ma, C.~Liang, R.~Yu, and S.~Liu, ``Isolation-improved
  dual-band mimo antenna array for lte/wimax mobile terminals,'' \emph{IEEE
  Antennas and wireless propagation letters}, vol.~13, pp. 1128--1131, 2014.

\bibitem{kwon2016integrated}
O.-Y. Kwon, R.~Song, Y.-Z. Ma, and B.-S. Kim, ``Integrated mimo antennas for
  lte and v2v applications,'' in \emph{2016 URSI Asia-Pacific Radio Science
  Conference (URSI AP-RASC)}.\hskip 1em plus 0.5em minus 0.4em\relax IEEE,
  2016, pp. 1057--1060.

\bibitem{bassi2013integrated}
M.~Bassi, M.~Caruso, M.~S. Khan, A.~Bevilacqua, A.-D. Capobianco, and
  A.~Neviani, ``An integrated microwave imaging radar with planar antennas for
  breast cancer detection,'' \emph{IEEE Transactions on microwave theory and
  techniques}, vol.~61, no.~5, pp. 2108--2118, 2013.

\end{thebibliography}

\begin{IEEEbiography}{M. S. Khan} is currently running a small company
related to RF projects. He received his B.Sc.
degree in electrical (telecom) engineering from
COMSATS University, Islamabad, Pakistan, in
2011. Based on his achievement during his B.Sc.
degree, he was awarded EMMA WEST Exchange
Scholarship for his B.Sc. mobility Program. He
received his Ph.D. degree in 2016 from University
of Padova, Italy. He was also recipient of
a fully funded Ph.D. scholarship from Cariparo
Foundation which provides scholarship to top 15 candidates from all over
the world. During his Ph.D., he also spent 18 months at North Dakota State
University, USA as a visiting scholar.

Dr. Saeed was Head of Department of Electrical Engineering Department
of Riphah International University, Lahore Campus for one year (2015-
2016). From 2016 to 2017, he also worked as a researcher (Post-doctorate
fellowship) with University of Padova. His current research interests include
advanced techniques and technologies for antenna design for medical
applications, phased array for radar systems, reconfigurable antennas for
advance applications, UWB-MIMO antennas, and novel material based
antennas. During this short period of time, he has authored/co-authored
more than 65 peer-reviewed journal or conference papers (h-index 14, more than 750 citations).
\end{IEEEbiography}

\begin{IEEEbiography}[{\includegraphics[width=1in,height=1.25in,clip,keepaspectratio]{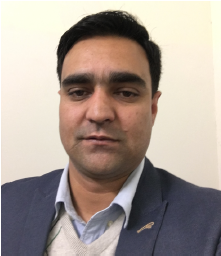}}]{A. Iftikhar} (S’12--M’16) received his B.S degree
in electrical engineering from COMSATS
University Islamabad (CUI), Pakistan, M.S degree
in Personal Mobile and Satellite Communication
from University of Bradford U.K., and
Ph.D. degree in Electrical and Computer Engineering
from North Dakota State University (NDSU), USA
in 2008, 2010, and 2016, respectively.

He has authored and co-authored 30 journals
and conference publications. His current research includes applied electromagnetic,
reconfigurable antennas, leaky wave antennas, phased array
antennas, UWB-MIMO antennas, and energy harvesting for low power devices.
He is member of IEEE and currently an Assistant Professor in Department of Electrical
and Computer Engineering at COMSATS University Islamabad (CUI),
Pakistan.
\end{IEEEbiography}

\begin{IEEEbiography}[{\includegraphics[width=1in,height=1.25in,clip,keepaspectratio]{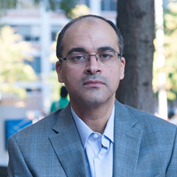}}]{Raed M. Shubair} (S’85–-M’93–-SM’01) is a Full Professor of Electrical Engineering and Visiting Scientist affiliated with MIT Research Laboratory of Electronics (RLE), Massachusetts Institute of Technology (MIT), USA and Department of Electrical and Computer Engineering, New York University (NYU), Abu Dhabi, UAE. Prof. Raed Shubair is also affiliated with Center of Intelligent Antennas and Radio Systems, Univer-sity of Waterloo, Canada and has been visiting at MIT Department of Brain and Cognitive Sciences and Harvard Medical School, USA.  Beside his academic positions at MIT, NYU, and Waterloo, he serves since 2017 as a Senior Advisor in the Office of Undersecretary for Academic Affairs of Higher Education, Ministry of Education, Abu Dhabi, UAE. He has been a Full Professor of Electrical Engineering at Khalifa University (formerly Etisalat University College), UAE, which he joined in 1993 up to 2017. Prof. Raed Shubair received his B.Sc. degree in Electrical Engineering (with Distinction and Class Honors) from Kuwait University, Kuwait in June 1989 followed by his Ph.D. degree in Electrical Engineering (with Distinction) from the University of Waterloo, Canada in February 1993. His PhD thesis received the University of Waterloo ‘Distinguished Doctorate Dissertation Award’. Prof. Raed Shubair has several research interests including Terahertz and Wireless Communications, Modern Antennas and Applied Electromagnetics, Signal Processing and Machine Learning, IoT and RF Localization, and Nano BioSensing.

Prof. Raed Shubair was the first faculty member in his university to be elevated to IEEE Senior Member grade in 2001  Prof. Raed Shubair has over 300 publications in the form of US patents, book chapters, and papers in IEEE transactions and IEEE conference proceedings. Prof. Raed Shubair received, several times since 1993, both the ‘University Teaching Excellence Award’ and ‘University Distinguished Service Award’. He is also recipient of several international awards including the ‘Distinguished Service Award’ from ACES Society, USA and from MIT Electromagnetics Academy, USA. Prof. Raed Shubair organized and chaired numerous technical special sessions and tutorials in IEEE flagship conferences. He delivered over 60 invited speaker seminars and technical talks in world-class universities and flagship conferences. Prof. Raed Shubair is a standing member of the editorial boards of several international journals and serves regularly on the steering, organizing, and technical committees of IEEE flagship conferences in Antennas, Communications, and Signal Processing including several editions of IEEE AP-S/URSI, EuCAP, IEEE GloablSIP, IEEE WCNC, and IEEE ICASSP. He has served as the TPC Chair of IEEE MMS2016 and TPC Chair of IEEE GlobalSIP 2018 Symposium on 5G Satellite Networks. Prof. Raed Shubair holds several leading roles in the international professional engineering community. He is Board Member of the European School of Antennas, Regional Director for IEEE Signal Processing Society in IEEE Region 8 Middle East, and Chair of IEEE Antennas and Propagation Society Educational Initiatives Program.  Prof. Raed Shubair is Fellow of MIT Electromagnetics Academy and Founding Member of MIT Scholars of the Emirates. Prof. Raed Shubair is Editor of the IEEE Journal of Electromagnetics, RF, and Microwaves in Medicine and Biology and Editor of IEEE Open Journal in Antennas and Propagation. Prof. Raed Shubair is a Founding Member of five IEEE society chapters in UAE which are IEEE-UAE Communication Society Chapter, IEEE-UAE Signal Processing Society Chapter, IEEE-UAE Antennas $\&$ Propagation Society Chapter, IEEE-UAE Microwave Theory and Techniques Society Chapter, IEEE-UAE Engineering in Medicine and Biology Society Chapter. Prof. Raed Shubair is the founder and counselor of the IEEE Student Branch at New York University Abu Dhabi.  Prof. Raed Shubair is a nominee for the Regional Director-at-Large of the IEEE Signal Processing Society in IEEE Region 8 Europe, Africa, and Middle East. Prof. Raed Shubair is also a nominee for the IEEE Distinguished Educator Award of the IEEE Antennas and Propagation Society. Prof. Raed Shubair has been honored and serves currently as an invited speaker with the prestigious U.S. National Academies of Sciences, Engineering, and Medicine.
\end{IEEEbiography}

\begin{IEEEbiography}[{\includegraphics[width=1in,height=1.25in,clip,keepaspectratio]{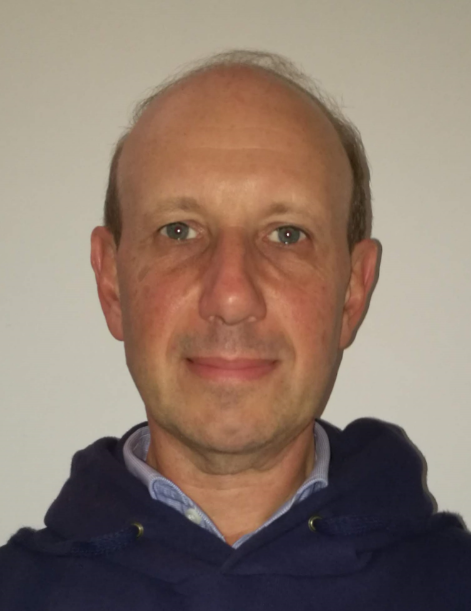}}]{Antonio-D. Capobianco} was born in Padova, Italy in 1965. He received the M.S. degree in electronic engineering from the University of Padova, in 1989 and the Ph.D. degree in electronic and communication engineering from Padova University, Italy, in 1994. From 1998 to 2017, he was an Assistant Professor with the Department of Information Engineering of Padova University. Since 2017, he has been an Associate Professor with the Department of Information Engineering of Padova University. He is the author of more than 160 articles. His research interests include silicon photonics, non-linear optics, nano antennas, microwaves antennas, and plasma antennas.
\end{IEEEbiography}

\begin{IEEEbiography}[{\includegraphics[width=1in,height=1.25in,clip,keepaspectratio]{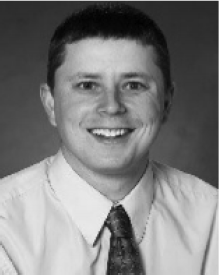}}]{Benjamin D. Braaten} (S’02–-M’09-–SM’14) received the B.S. degree in electrical engineering
in 2002, the M.S. degree in electrical engineering
in 2005, and the Ph.D. degree in electrical and
computer engineering in 2009, all from the North
Dakota State University, Fargo, ND, USA.

During the 2009 Fall semester, he held a Postdoctoral
Research Position in the South Dakota
School of Mines and Technology, Rapid City, SD,
USA. At the end of the 2009 Fall Semester, he
joined the Faculty of the Electrical and Computer Engineering Department,
North Dakota State University, and was promoted to Associate Professor
with tenure in 2015. He has authored or co-authored more than 100
peer reviewed journal and conference publications, several book chapters
on the design of antennas for radio frequency identification, and holds one
U.S. patent on wireless pacing of the human heart. His research interests
include printed antennas, conformal self-adapting antennas, microwave
devices, topics in EMC, topics in BIO EM, and methods in computational
electromagnetics.

Dr. Braaten received the College of Engineering and Architecture Graduate
Researcher of the Year and College of Engineering and Architecture
Graduate Teacher of the Year awards. He also serves as an Associate Editor
for the IEEE ANTENNAS AND WIRELESS PROPAGATION LETTERS
and is a member of the National Honorary Mathematical Society PI MU
EPSILON. Currently, Prof. Dr. Braaten is chairman of ECE department at
NDSU, Fargo, ND, USA.
\end{IEEEbiography}

\begin{IEEEbiography}[{\includegraphics[width=1in,height=1.25in,clip,keepaspectratio]{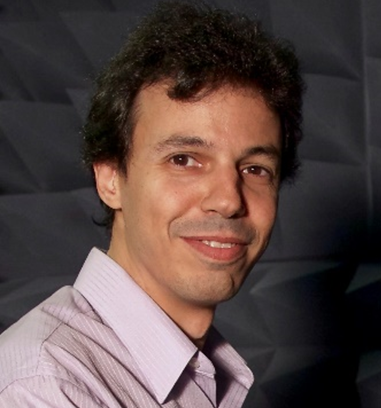}}]{Dimitris E. Anagnostou} (S’98--M’05-–SM’10) received the B.S.E.E. degree from the Democritus University of Thrace, Greece, in 2000, and the M.S.E.E. and Ph.D. degrees from the University of New Mexico, Albuquerque, NM, in 2002 and 2005, respectively. From 2005 to 2006, he was a Postdoctoral Fellow with the Georgia Institute of Technology, Atlanta, GA. In 2007, he joined as Assistant Professor the SD School of Mines $\&$ Technology, SD, USA, where he was promoted to Associate Professor with tenure. In 2016, he joined the Heriot Watt University, Institute of Signals, Sensors and Systems (ISSS), Edinburgh, UK as an Associate Professor. He has also worked at the Kirtland AFB, NM in 2011 as an AFRL Summer Faculty Fellow, and at the Democritus Univ. of Thrace, Greece as Assistant Professor.

Dr. Anagnostou is currently supported by a European H2020 Marie Skłodowska-Curie Individual Fellowship on wireless sensing technologies. He has authored or coauthored more than 150 peer-reviewed journal and conference publications (h-index 24, more than 2200 citations), one book chapter, and holds two U.S. patents on MEMS antennas and on optically scannable QR-code ‘anti-counterfeiting’ antennas. His research interests include: antennas (reconfigurable, adaptive, miniaturized and electrically small, space/satellite, wearable), microwave circuits and packaging, radar sensing, 5G arrays, functional phase-change materials such as VO2 for reconfigurable electronics and metasurfaces, direct-write electronics on organics, RF-MEMS, and applications of artificial neural networks, deep learning and signal processing in electromagnetics, health care and assisted living.

Dr. Anagnostou serves or has served as Associate Editor for the \emph{TRANSACTIONS ON ANTENNAS AND PROPAGATION} (2010-2016) and the \emph{IET MICROWAVES, ANTENNAS AND PROPAGATION} (since 2015). He is Guest Editor He is Guest Editor for \emph{IEEE ANTENNAS AND WIRELESS PROPAGATION LETTERS} (two Special Clusters), and \emph{MDPI Electronic}. He is a member of the \emph{IEEE AP-S Educational Committee}, and of the Technical Program Committee (TPC) of the \emph{IEEE AP-S} and \emph{EuCAP International Symposia}. He has received the 2010 \emph{IEEE John D. Kraus Antenna Award}, the 2011 \emph{DARPA Young Faculty Award} by the U.S. Department of Defense, the 2014 \emph{Campus Star Award} by the American Society for Engineering Education (ASEE), the 2017 \emph{Young Alumni Award} by the University of New Mexico, the \emph{H2020 MSCA Fellowship}, and four \emph{Honored Faculty Awards} by SDSMT. His students have also been recognized with IEEE and university awards (\emph{Engineering Prize} at Heriot Watt University, \emph{Best PhD Thesis} at South Dakota School of Mines, and others). He is a member of HKN Honor Society, ASEE, and of the Technical Chamber of Greece as a registered Professional Engineer (PE).
\end{IEEEbiography}

\end{document}